\documentclass[twocolumn,showpacs,preprintnumbers,amsmath,amssymb]{revtex4}
\usepackage{graphicx}
\begin{document}
\title{Kondo effects in small bandgap carbon nanotube quantum dots}
\author{P. Flork\'{o}w, D. Krychowski and S. Lipi\'nski}
\affiliation{%
Institute of Molecular Physics, Polish Academy of Sciences\\M. Smoluchowskiego 17,
60-179 Pozna\'{n}, Poland
}%
\date{\today}
\begin{abstract}
We study magnetoconductance of  the small bandgap carbon nanotube quantum dots in the presence of spin-orbit coupling in the strong correlations regime. The finite-U mean field slave boson approach is used to study many-body effects.  Different degeneracies are restored in magnetic field  and Kondo effects of different symmetries arise including SU(3) effects of different types.  Full spin-orbital degeneracy might be recovered for zero field and correspondingly  SU(4) Kondo effect sets in. We point out on the possibility of the occurrence of electron-hole Kondo effects in slanting magnetic fields, which we predict will occur in the available  magnetic fields for orientation of fields close to perpendicular. When the field approaches transverse orientation a crossover from SU(2) or SU(3) symmetry into SU(4) is observed.
\end{abstract}

\maketitle

\section{Introduction}

Due to remarkable electronic, transport, mechanical and magnetic properties carbon nanotubes (CNTs) are of great interest in molecular electronics and spintronics with potential applications e.g. as field effect transistors, nanoelectromechanical devices, logic gates, spin valves, spin diodes, spin batteries \cite{McEuen,Avouris1,Dresselhaus,Avouris2,Saito,Cottet,Javey,Shulaker,Laird,Lipinski,Merchant}. CNTs are also interesting for fundamental science. Their study allows examination of many basic properties in the ranges often not reachable in other systems. Many of the fundamental transport properties were observed in nanotubes including Coulomb blockade \cite{Nersesyan,Grove}, Fabry-Perot interference \cite{Grove,Peca}, Kondo physics \cite{Jarillo,Makarovski,Zhukov,Wu,Choi1,Lim,Martins,Fang1,Anders,Mizumo,Lipinski2,Krychowski,Mantelli,Galpin,Schmid}, spintronic effects \cite{Lipinski,Kuemmeth1,Zhao,Tombros} and induced superconductivity \cite{Ferrier,Sasaki,Tang}.

In the present paper we are interested in the effects of strong correlations. As electrons are confined in fewer dimensions, the effects of interactions play a more fundamental role. Carbon nanotubes are quasi one-dimensional systems, and the role of correlations further increases in quantum dots due to additional confinement. Of importance is also low dielectric constant, especially low in suspended nanotubes \cite{Cao}.

Single wall carbon nanotube is a hollow cylinder formed from graphene. CNT can be either metallic or semiconducting depending on the way graphene is rolled up \cite{Odom,Wilder}. In the simple \textit{zone folding} picture \cite{Hamada,Sato} the band structure of CNT  is obtained  from the band structure of graphene by imposing periodic conditions along circumference. When the closest quantization line misses the K point a band gap appears. Structural bandgap depends on the minimum separation of the circular quantization lines from Dirac points. The semiconducting gaps are of  order of few hundred meV \cite{Laird}. The nanotubes are characterized by chiral vector C(n,m), where the integers $n$ and $m$ denote the number of unit vectors along two directions in the honeycomb crystal lattice of graphene. Zone folding theory predicts nanotube to be metallic if n-m is a multiple of $3$. According to this prediction one third of  randomly selected nanotubes should be metallic.

Experimentally, the fraction of the nanotubes showing metallic behavior is very small ($\leq 1\%$ \cite{Churchill}).  Even in nominally metallic tubes the narrow gap of order of $10$ meV is usually observed. These systems are sometimes called nearly metallic carbon nanotubes \cite{Steele1,Maslyuk}. The reason for incorrect predictions of zone folding theory for small diameter tubes is  neglect of   the  curvature induced breaking of the  three-fold ($C_{3}$)   rotational symmetry. When graphene is deformed into a nanotube, the curvature causes the overlap matrix elements to depend on direction \cite{Kane}. The consequence of breaking symmetry is a shift of  Dirac points in reciprocal lattice  away from $K$ and $K'$ points. The broken symmetry also enhances the intrinsic spin-orbit (SO) coupling in carbon nanotubes compared with the flat graphene. Apart from curvature \cite{Kane,Izumida,Park,Yang} also other perturbations like axial strain or twists can shift the dispersion cones in CNTs and open the gap \cite{Kane,Yang1,Minot1,Rochefort}. Unlike the quantization bandgaps, which depend on the inverse of diameter, these narrow perturbation gaps are inversely proportional to the square of diameter and depend on the  chiral angle. Small gaps are reflected in nonlinear dispersion curves and consequent drastic changes of orbital effects induced by magnetic field. The behavior in magnetic fields is distinctly different than in wide gap nanotubes. The field dependencies are determined not only by the response of orbital and spin magnetic moments, as in the case of large gaps, but crucially  depend also on the value of the gap and  gate voltage. Details of the band structure are decisive for the response on the field. The degeneracy recovery lines placed  on the plane of  magnetic field and gate voltage are no longer parallel to the gate voltage axis, like in the wide gap tubes, but they are gate voltage dependent  and they  intersect at certain fields with other similar lines, what means the appearance of higher degeneracy in the system. The presence of different degeneracy points and the gate dependence of degeneracy lines  is interesting for quantum computing, because it opens the possibility of electric switching between different types of qubits (spin, valley or valley-spin) and their higher dimensional equivalents (qutrits\cite{Li,Lanyon}, qudits \cite{Luo,Pineda}) at the same nanoscopic system. Storage capacities of  the three-state or four-state qudits are obviously higher than capacity of qubit. In the present paper we are interested in transport properties. Transmission of the contacts of the quantum dot with electrodes determines the regime of charge transport. For very weak transparency charging effects dominate transport at low temperatures and the electrons enter the dot one by one yielding the well known Coulomb blockade oscillations as a function of gate voltage. For more open contacts the role of higher order tunneling processes (cotunneling) increases, what at low temperatures results in formation of many-body resonances at the Fermi level, and the  new transport paths open in the valleys between Coulomb peaks. In contrast to graphene, where spin-orbit interaction is weak due to inversion symmetry of graphene plane, in CNTs this symmetry is broken due to curvature and in consequence the hopping between $p$ orbitals of different parity from different atoms is allowed what leads to substantial SO coupling. The single electron spectroscopy measurements done on ultraclean nanotubes showed that even at zero magnetic field the spin and orbital degrees of freedom are not independent and a level splitting into two Kramers doublets has been observed \cite{Kuemmeth}. This effect has been ascribed to spin-orbit interaction. Later many other experiments confirmed the importance of SO coupling in CNTs \cite{Steele,Grove2,Jhang,Pei,Cleziou}. Depending on the sign of SO coupling this interaction introduces parallel or antiparallel alignment of spin and angular momentum. The energy of SO coupling is comparable to  the energy scale of Kondo effect and therefore taking this perturbation  into account is important when analyzing many-body effects in these systems. Several interesting papers has been devoted to the problem of interplay of Kondo effect and SO interaction \cite{Krychowski,Schmid,Fang,Mantelli,Niklas}.	
Apart from SU(2) Kondo resonances  with effective  spin, valley or spin-valley fluctuations  high degeneration points  allow also the occurrence of  many body resonances of SU(3) and SU(4) symmetries. In the following we focus on the description of these exotic Kondo effects influenced by a subtle interplay of magnetic field, spin-orbit interaction and changes of the bandgap. In particular, we show how for a given nearly metallic  nanotube  one can change the position of high symmetry points by strain and magnetic field. Our calculations also show that in quantum dot formed in small gap nanotube  electron and hole states can degenerate in the slanting magnetic fields. Based on this observation we anticipate the possibility of occurrence of Kondo effects in which both these types of carriers take part. Apart from SU(2) Kondo lines also SU(3) Kondo points and SU(4) may appear for orientations of the field close to perpendicular.

\section{Model and formalism}
In our analysis we consider the low energy and low temperature range and therefore we restrict for the most part of our discussion to  only single shell of  carbon nanotube energy states, it is to four states labeled by spin ($s =\pm1$) and valley pseudospin ($l =\pm1$).  The model we use to describe CNTQD is extended two-orbital Anderson model:
\begin{eqnarray}
&&{\mathcal{H}}={\mathcal{H}}_{d}+{\mathcal{H}}_{L}+{\mathcal{H}}_{R}+{\mathcal{H}}_{t}
\end{eqnarray}
where the dot Hamilonian reads:
\begin{eqnarray}
&&{\mathcal{H}}_{d}=\sum_{ls}E^{e(h)}_{ls}N_{ls}+U\sum_{lss'}(N_{l\uparrow}N_{l\downarrow}
+N_{1s}N_{-1s'})
\end{eqnarray}
with site dot energies:
\begin{eqnarray}
&&E^{e(h)}_{ls}=\pm \sqrt{(l\mu_{o}B_{\|}+ls\Delta_{O}+E_{g})^{2}+E_{d}(V_{g})^{2}}
\nonumber\\&&+ls\Delta_{Z}+s\frac{g\mu_{B}B_{\|}}{2}
\end{eqnarray}
dependent on magnetic field $B_{\|}$ and gate voltage $V_{g}$. The upper and lower signs $\pm$ refer to conduction or valence states, $\mu_{o}$  is orbital magnetic moment $\mu_{o} = ev_{F}D/4$, where $v_{F}$  is the Fermi velocity ($v_{F}\cong0.8c$),  $D$ is nanotube diameter $D(n,m)=\frac{a}{\pi}\sqrt{m^{2}+mn+n^{2}}$ and  $a$ is the distance  between carbon atoms in A (B) lattice of graphene ($a\cong0.254$ nm). $E_{g}$ is the bandgap $E_{g}=\beta\frac{cos(3\Theta)}{D^{2}}$, where $\Theta$ is chiral angle ($\Theta=atan(\sqrt{3}m/(2n+m))$). According to tight-binding calculations the value of $\beta$ corresponding to the equilibrium energy gap is equal $37$ meV $nm^{2}$ \cite{Laird,Jespersen}. $\Delta_{O}$, $\Delta_{Z}$ stand for orbital and Zeeman parameters of spin-orbit coupling  taken in the form:
\begin{eqnarray}
&&{\mathcal{H}}_{SO}=\Delta_{O}s_{z}l_{x}+\Delta_{Z}ls_{z}
\end{eqnarray}
where $s_{z}$ is the spin component along the nanotube axis and $l_{x}$ is Pauli matrix in the A-B graphene sublattice space. $\Delta_{Z}=-\delta/D$ and $\Delta_{O}=\delta cos(3\Theta)/D$ \cite{Jespersen}. Various theoretical and experimental estimates differ not only in the reported  values of parameter $\delta$, but also often in predictions of its sign. It ranges from  one tenth  to few meV nm \cite{Steele}. For wide bandgap nanotubes  $E_{g}\gg\Delta_{Z}, \Delta_{o}$ the field dependence of single particle energies (3) becomes linear ($E^{e(h)}_{ls}\approx \pm\sqrt{E_{g}^{2}+E_{\|}^{2}}+ls\frac{\Delta_{e(h)}}{2}+\frac{sg\mu_{B}\mp l\mu_{o}}{2}B_{\|}$), for small gap it is parabolic. In large gap case SO splitting can be described by one effective parameter $\Delta^{e(h)} = (\Delta_{Z}\mp\frac{\Delta_{O}}{\sqrt{1+(E_{\|}/E_{g})^{2}}})$.

${\cal{H}}_{L}+{\cal{H}}_{R}$ describe electrons in the left and right electrodes:
\begin{eqnarray}
&&{\mathcal{H}}_{\alpha}=\sum_{k\alpha ls}E_{k\alpha ls} c^{\dag}_{k\alpha ls}c_{k\alpha ls}
\end{eqnarray}
($\alpha = L,R$) and the last term in (1) represents tunneling:
\begin{eqnarray}
&&{\mathcal{H}}_{t}=\sum_{k\alpha ls}t(c^{\dag}_{k\alpha ls}d_{ls}+h.c.)
\end{eqnarray}
We parametrize coupling strength to the leads by $\Gamma = \sum_{\alpha}\Gamma_{\alpha}=\sum_{k\alpha ls}\pi t^{2}\varrho_{\alpha ls}$. In the following the wide conduction band approximation with the rectangular density of states is used $\varrho_{\alpha ls}(E)=1/2W$, where $W$ is the half-bandwidth.

To analyze correlation effects we use slave boson mean field approach of Kotliar and Ruckenstein (K-R) \cite{Krychowski,Kotliar,Dong}. In this picture different auxiliary bosons are introduced to project onto different orbital and spin states. Apart from empty state boson $e$, single occupied $p_{ls}$ and double occupied $d$ also triply occupied $t_{ls}$ and fully occupied $f$ are used. The $p$ operators are labeled by indices specifying the corresponding single-particle states, $t$ bosons  are characterized by hole indexes and six $d$ operators project onto doubly occupied states $d_{l=\pm 1}$:
$|20\rangle$, $|02\rangle$ and $d_{ss'}$: $|\uparrow\uparrow\rangle$, $|\downarrow\downarrow\rangle$,
$|\uparrow\downarrow\rangle$ and $|\downarrow\uparrow\rangle$. To
eliminate unphysical states, the completeness relations
for the slave boson operators (${\cal{I}}=e^{\dag}e+\sum_{lss'}(p^{\dag}_{ls}p_{ls}
+d^{\dag}_{l}d_{l}+d^{\dag}_{ss'}d_{ss'}
+t^{\dag}_{ls}t_{ls})+f^{\dag}f$) and the conditions for the
correspondence ($Q_{ls} = p^{\dag}_{ls}p_{ls}+d^{\dag}_{l}d_{l}+d^{\dag}_{ss}d_{ss}
+d^{\dag}_{s\overline{s}}d_{s\overline{s}}+t^{\dag}_{ls}t_{ls}
+t^{\dag}_{\overline{l}s}t_{\overline{l}s}+t^{\dag}_{\overline{l}\overline{s}}t_{\overline{l}\overline{s}}
+f^{\dag}f$) between fermions and bosons have to be
imposed. These constraints can be enforced by introducing Lagrange multipliers $\lambda$ and $\lambda_{ls}$.
In K-R approach Hamiltonian (1) describing interacting fermions is replaced by an effective Hamiltonian  of noninteracting bosons and pseudofermions. It takes the form:
\begin{eqnarray}
&&{\mathcal{\widetilde{H}}}=\sum_{k\alpha ls}E^{e(h)}_{ls}N^{f}_{ls}+\sum_{lss'}U(d^{\dag}_{l}d_{l}+d^{\dag}_{ss'}d_{ss'})+\nonumber\\
&&+\sum_{ls}3Ut^{\dag}_{ls}t_{ls}+6Uf^{\dag}f+\sum_{ls}(\lambda_{ls}-Q_{ls})
+\lambda(I-1)\nonumber\\
&&+\sum_{k\alpha ls}E_{k\alpha ls}N_{k\alpha ls}+\sum_{k\alpha ls}t(c^{\dag}_{k\alpha ls}z_{ls}f_{ls}+h.c.)
\end{eqnarray}
where $N^{f}_{ls}=f^{\dag}_{ls}f_{ls}$ are the pseudofermion occupation operators and $f_{ls}$ is defined by $f_{ls}=d_{ls}z_{ls}$ with boson operator $z_{ls}=(e^{\dag}p_{ls}+p^{\dag}_{l\overline{s}}d_{l}
+p^{\dag}_{\overline{l}\overline{s}}(\delta_{l,1}d_{s\overline{s}}+\delta_{l,-1}d_{\overline{s}s})
+p^{\dag}_{\overline{l}s}d_{ss}+
d^{\dag}_{\overline{l}}t_{ls}+d^{\dag}_{\overline{s}\overline{s}}t_{\overline{l}\overline{s}}
+(\delta_{l,-1}d^{\dag}_{s\overline{s}}+\delta_{l,1}d^{\dag}_{\overline{s}s})t_{\overline{l}s}
+t^{\dag}_{l\overline{s}}f)/\sqrt{Q_{ls}(1-Q_{ls})}$.
The mean field solutions are found from the minimum of the free energy with respect to the mean values of slave boson operators and Lagrange multipliers. SBMFA method is correct in the unitary Kondo regime and it leads to a local Fermi liquid behavior at zero temperature. It gives reliable results of linear conductance also for systems with weakly broken symmetry. The results obtained are in good agreement with experiment and with  the renormalization group calculations \cite{Mantelli,Galpin}. SBMFA method  works worse at higher temperatures and therefore if we analyze temperature dependencies, we use a complementary approach of equation of motion method (EOM) with Lacroix approximation \cite{Lacroix}, which approximates the Green functions involving two conduction- electron operators by:
\begin{eqnarray}
\langle\langle c^{\dag}_{k\alpha l's'}d_{l's'}c_{k\alpha ls};d^{\dag}_{ls}\rangle\rangle\simeq\langle c^{\dag}_{k\alpha l's'}d_{l's'}\rangle\langle\langle c_{k\alpha ls};d^{\dag}_{ls}\rangle\rangle\\
\langle\langle c^{\dag}_{k\alpha l's'}c_{k\alpha l's'}d_{ls};d^{\dag}_{ls}\rangle\rangle\simeq\langle c^{\dag}_{k\alpha l's'}c_{k\alpha l's'}\rangle\langle\langle d_{ls};d^{\dag}_{ls}\rangle\rangle\nonumber
\end{eqnarray}
The correlations $\langle c^{\dag}_{k\alpha l's'}d_{l's'}\rangle$ and $\langle c^{\dag}_{k\alpha l's'}c_{k\alpha l's'}\rangle$  play the leading role in Kondo effect. The advantage of EOM method in comparison to SBMFA is that EOM accounts not only for spin or pseudospin fluctuations, but also for charge fluctuations and allows  to obtain  the information about the behavior of the system also at higher temperatures.

Physical quantities that are the object of our interest are linear conductance  ${\cal{G}}$ and thermoelectric power ${\cal{S}}$. Both of these quantities can be determined from the transmissions, which in turn can be calculated from the knowledge of Green's functions obtained in SBMFA or EOM. ${\cal{G}}=\sum_{ls}{\cal{G}}_{ls}=\sum_{ls}(e^{2}/h)L_{ls,0}/T$, ${\cal{S}}=\sum_{ls}(-k_{B}/e)L_{ls,1}/(T\sum_{ls}L_{ls,0})$, where $L_{ls,n}=\sum_{\alpha}\int^{+\infty}_{-\infty} (E-\mu_{\alpha})^{n}f_{\alpha}(E){\cal{T}}_{ls}(E)dE$.  $f_{\alpha}(E)$ are the Fermi distribution function of electrodes and $\mu_{\alpha}=\pm V_{sd}/2$. ${\cal{T}}_{ls}$ is the spin-orbital transmission. For spintronics and valleytronics important quantities are spin (SPC) or orbital (OPC) polarizations of conductance directly expressed through partial conductances SPC$=(\sum_{l}{\cal{G}}_{l\uparrow}-{\cal{G}}_{l\downarrow})/{\cal{G}}$, OPC$=(\sum_{s}{\cal{G}}_{1s}-{\cal{G}}_{-1s})/{\cal{G}}$. Spin and orbital magnetic moments are defined as $M_{Z}=\sum_{l}(N_{l\uparrow}-N_{l\downarrow})$ and $T_{z}=\sum_{s}(N_{1s}-N_{-1s})$, where $N_{ls}$ denote electron occupations.

\section{Results}
\begin{figure}
\includegraphics[width=0.48\linewidth]{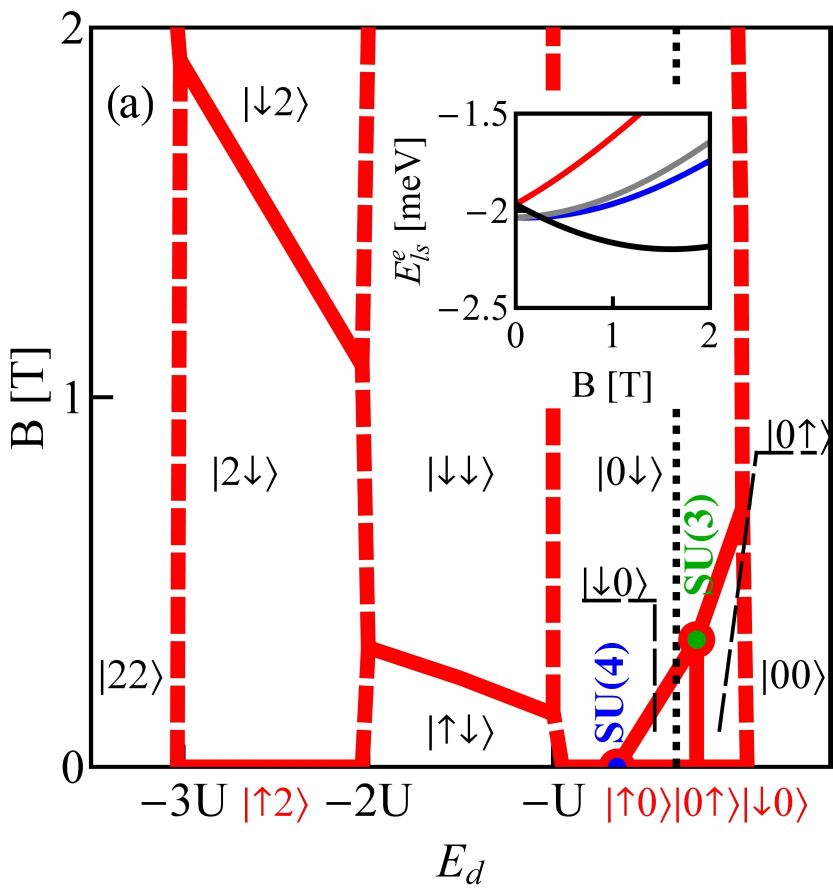}
\includegraphics[width=0.48\linewidth]{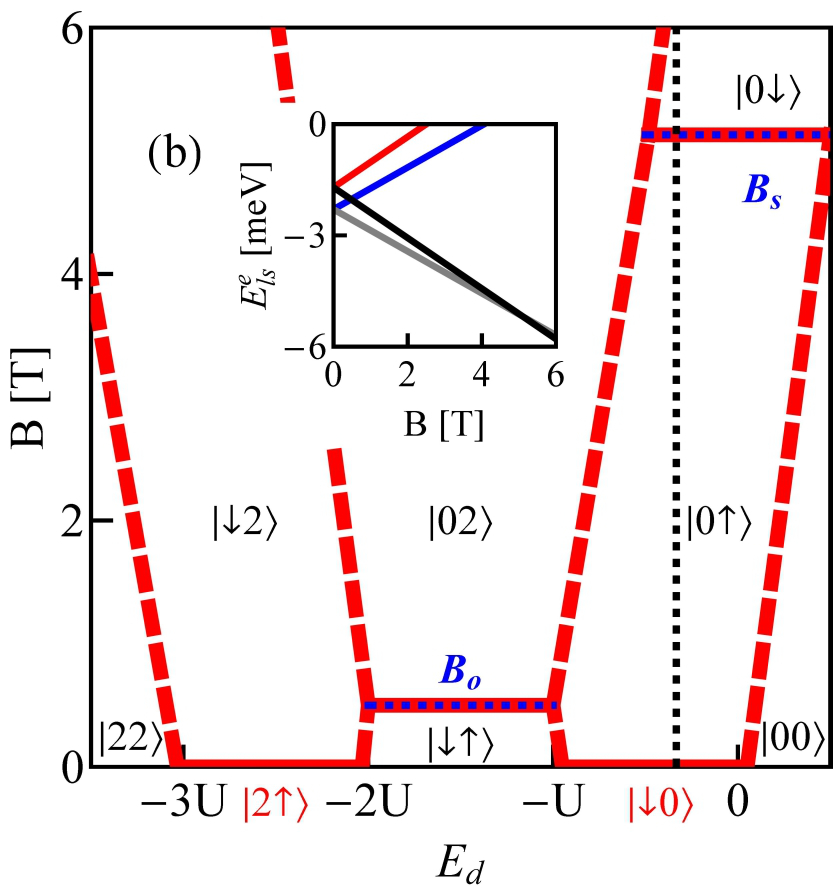}\\
\includegraphics[width=0.48\linewidth]{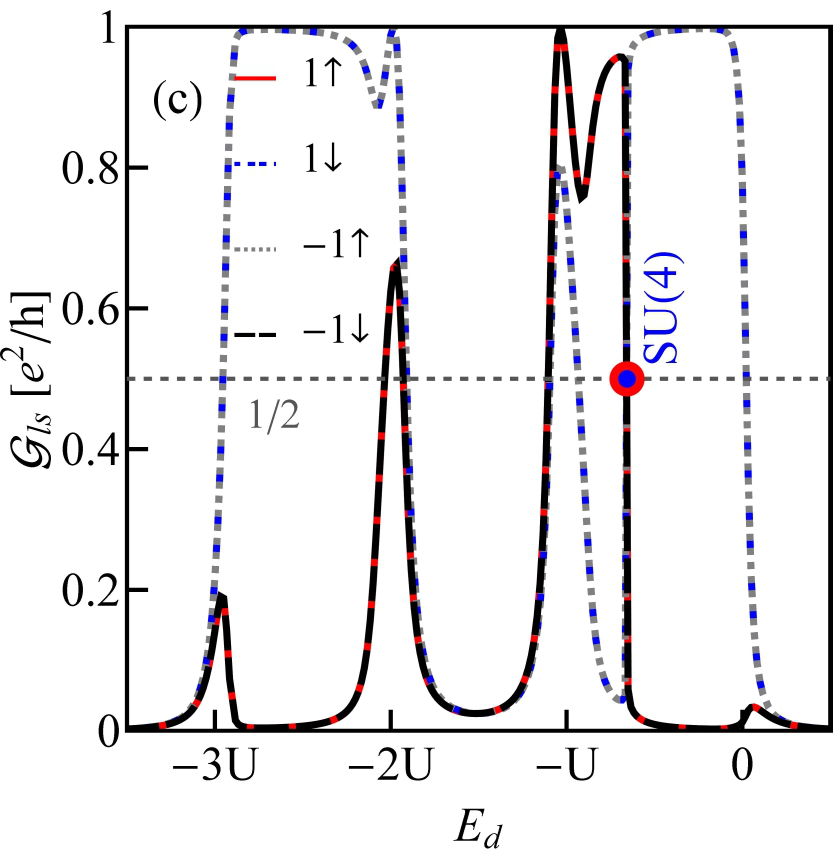}
\includegraphics[width=0.48\linewidth]{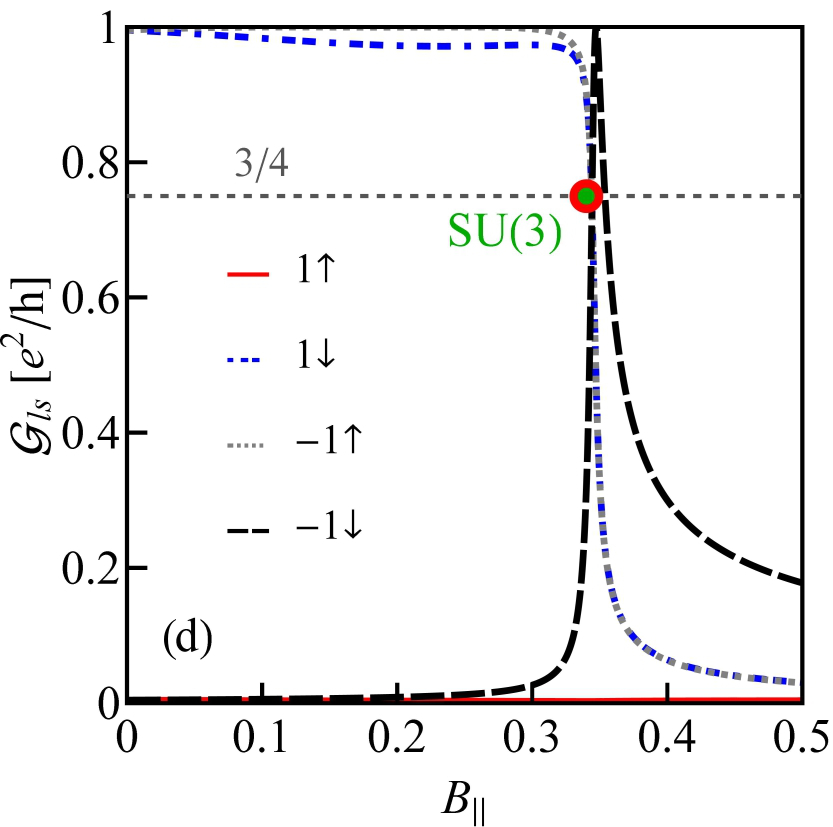}
\caption{\label{fig:fig1} (a),(b) Ground state maps of isolated quantum dots ($U = 6$ meV, $\delta=1$ meV nm): (a) dot formed in small gap nanotube C(24,21) $E_{g}=0.45$ meV (this gap corresponds to $\beta=37$ meV $nm^{2}$), (b) dot in semiconducting tube C(24,22) ($E_{g}=125$ meV). The broken red lines are the borders between Coulomb blockade valleys and solid lines are double degeneracy lines. Blue and green points denote triple and fourfold degeneracy points respectively. The black brackets $|ls\rangle$ mean the ground states for $B>0$ and brackets in red are the ground states for $B<0$. Insets show magnetic field dependencies of energies of single particle states $|\uparrow0\rangle$, $|\downarrow0\rangle$, $|0\uparrow\rangle$ and $|0\downarrow\rangle$ (red, blue, gray and black lines) plotted for the cross-sections denoted by dotted black vertical lines. (c) partial conductances of the dot CNTQD(24,21) strongly coupled to the leads for $B_{\|}=0$ (d) conductances for fixed values of gate voltage (cross-section through SU(3) point specified by dotted black line) ($\Gamma=0.03$ meV).}
\end{figure}

All the numerical results presented below concern quantum dots formed in nearly metallic  nanotubes with perturbation gaps. We compare in figure \ref{fig:fig1} the ground states diagram of isolated quantum dot formed in  small gap nanotube C(24,21)($n = 24$, $m = 21$, band gap $E_{g} = 0.46$ meV, figure \ref{fig:fig1}(a)) with the diagram for the dot in wide gap tube C(24,22) (band gap $E_{g} = 125$ meV, figure \ref{fig:fig1}(b)). Insets present field dependencies of single electron states, which according to formula (2) are linear for the wide gap tubes and parabolic for tubes with narrow gaps. For vanishing magnetic field the ground states in odd Coulomb valleys are  degenerate (Kramers degeneracy). For the assumed SO  parameters ($\Delta_{Z}=-0.02$ meV, $\Delta_{O}=-0.32$ meV) ($\beta=37$ meV $nm^{2}$). The ground state doublet is  $\{|0\uparrow\rangle,|\downarrow0\rangle\}$ in 1e valley and $\{|\downarrow2\rangle,|2\uparrow\rangle\}$ in 3e valley. In  even valley the ground state is singlet $|\downarrow\uparrow\rangle$. Magnetic field breaks time-inversion symmetry what results in splitting of Kramers doublets in odd valleys. Depending on the signs of SO coupling the recovery of degeneracy resulting from a competition of  Zeeman effect and SO interaction can result in 1e or 3e valley \cite{Laird}. For the analyzed example of wide gap tube a crossing of energy levels occurs in 1e valley. For the field $B_{s} = |\Delta_{e}|/g\mu_{B} \approx 5.17$ T, the energy of the state $|0\uparrow\rangle$ is crossed by energy line of one of the states from higher Kramers doublet  $|0\downarrow\rangle$ (inset on figure \ref{fig:fig1}(b)) and degeneracy is recovered. The characteristic field is determined by SO splitting alone and therefore the degeneracy line is parallel to the gate voltage axis. Recovery of degeneracy is also observed in 2e valley, two states $|\downarrow\uparrow\rangle$ and $|02\rangle$ degenerate in magnetic field $B_{o} = |\Delta_{e}|/\mu_{o} \approx 0.48$ T. In small gap nanotubes the field dependencies of degeneracy lines are determined not only by spin-orbit parameters, orbital and spin magnetic moments, but also by the gap and gate voltage. This is reflected in the  nonlinear gate dependencies of degeneracy lines and the possibility of degeneration of more than two states. The fact that the boundaries between areas of different ground states are not parallel to the gate  axis opens the path for electric control of transitions between different ground states, and consequently it also enables switching  of such physical quantities as e.g. magnetic or orbital  moments of the  dots (examples of the  maps of magnetic and orbital moments are given on figures \ref{fig:fig3}(c), (d)). The occurrence of points of different degeneration and the ability to move between them by change of the gate voltage may have significance for quantum computing, because it provides  a method for electric switching  in the same nanoscopic system between spin, valley or spin valley  qubits, as well as between qubits and qutrites of various types (triple degeneration), or qudits (fourfold degeneration). For the analyzed nanotube C(24,21)  threefold and fourfold degenerate  points appear in 1e valley. SU(3) point  is observed for finite field in the single occupancy region  and fourfold occurs in the same valley for zero field.
\begin{figure}
\includegraphics[width=0.48\linewidth]{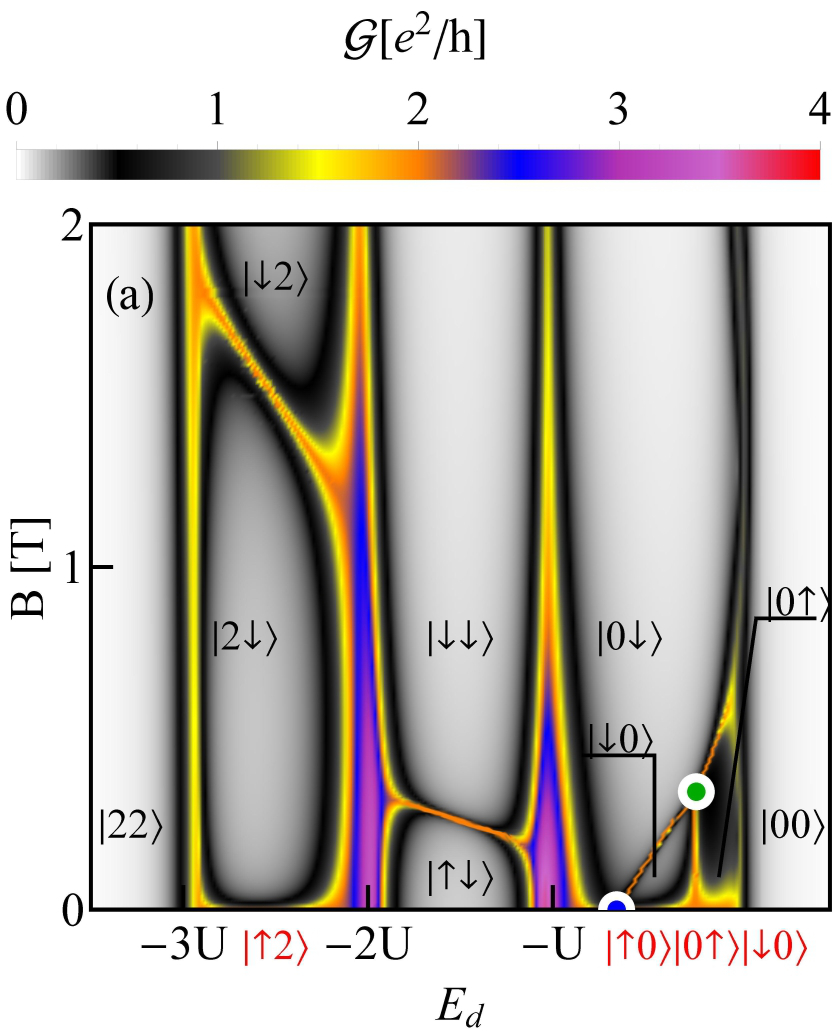}
\includegraphics[width=0.48\linewidth]{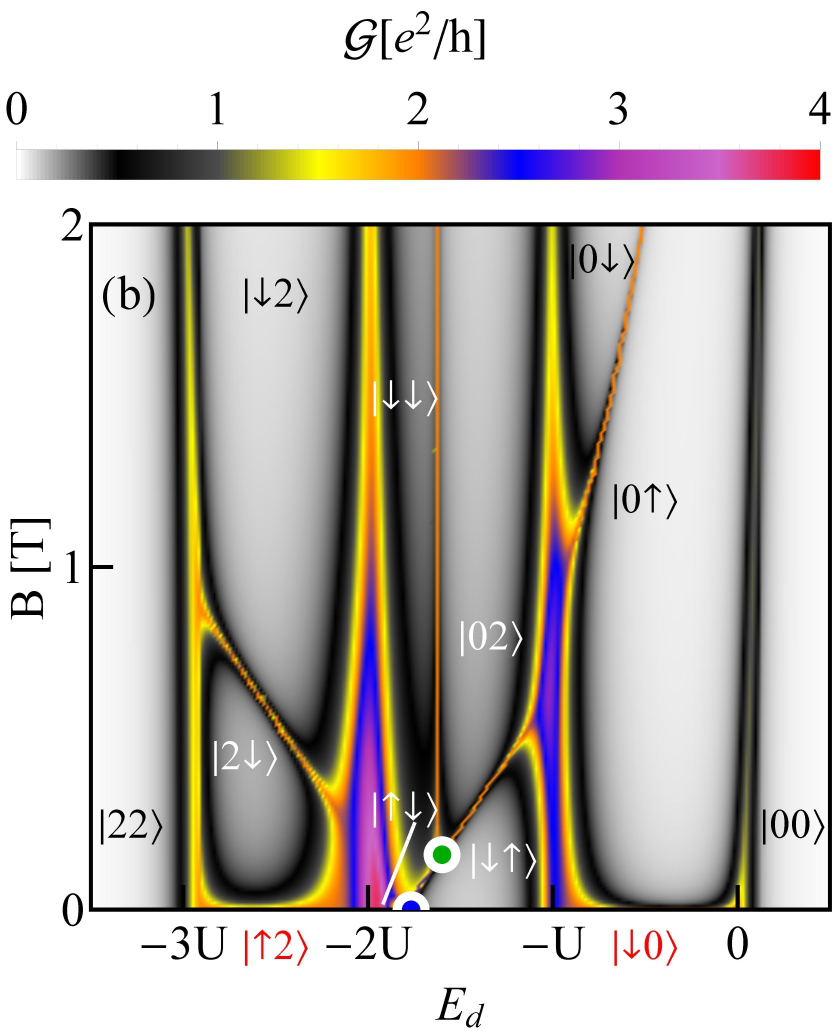}\\
\includegraphics[width=0.48\linewidth]{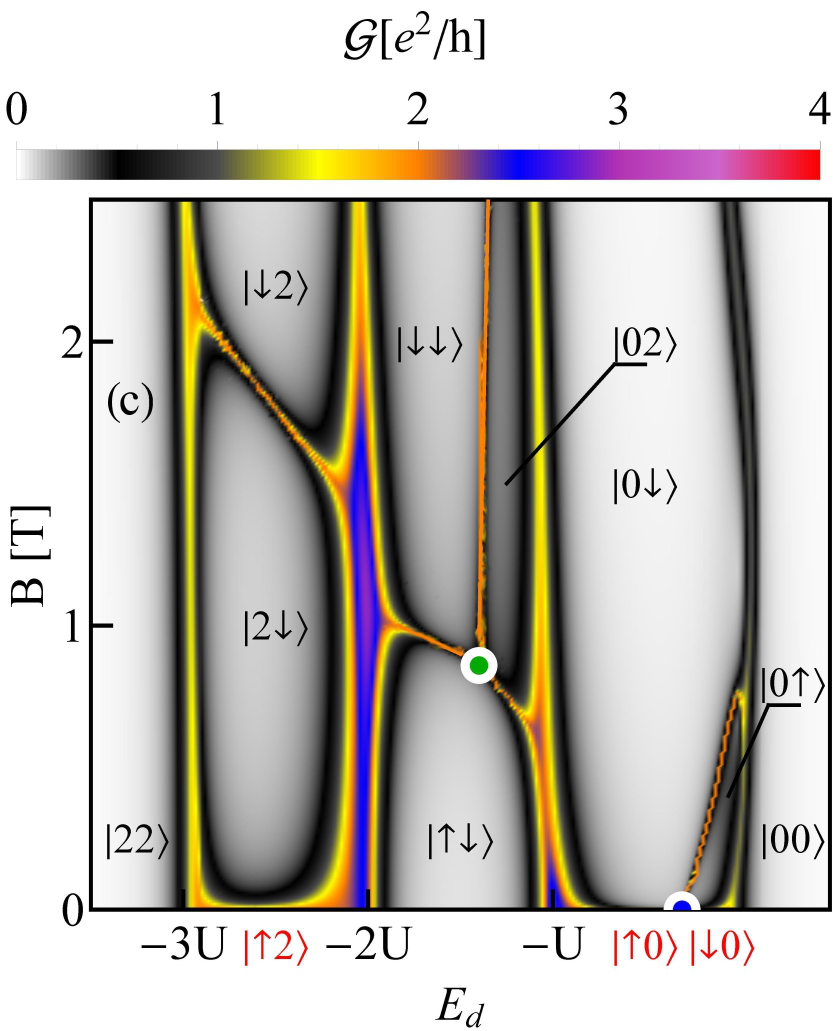}
\includegraphics[width=0.48\linewidth]{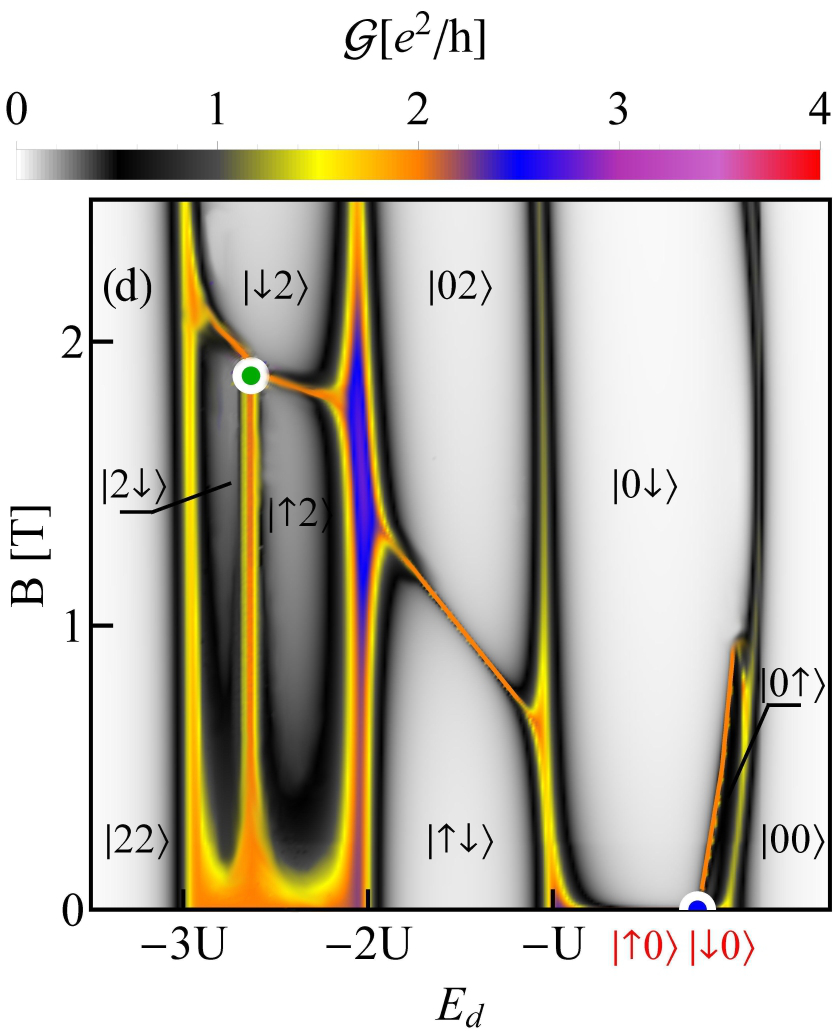}
\caption{\label{fig:fig2} Maps of the total conductances with plotted ground state diagrams of (a) CNTQD(24,21), (b) CNTQD(39,24), (c) CNTQD(15,12) and (d) CNTQD(48,18) ($U=6$ meV, $\Gamma=0.03$ meV, $\delta=1$ meV nm, $\beta=37$ meV $nm^{2}$).}
\end{figure}

Up till now we described the properties  of isolated carbon nanotube quantum dots (CNTQDs)  and now let us concentrate on their  transport properties in the strong correlation regime. Figures \ref{fig:fig1}(c), (d) present partial conductances of the dot formed in nanotube C(24,21): CNTQD(24,21) for the cases when the dot is strongly coupled to the leads and Kondo effects  occur at the degeneracy points or lines. Figure \ref{fig:fig1}(c) presents gate dependencies of partial conductances for zero magnetic field,  where SU(4) spin–orbital Kondo point separates intervals of occurrence of SU(2) Kondo effects related to two different Kramers doublets. For lower gate voltages   $\{|\uparrow0\rangle,|0\downarrow\rangle\}$  states are active in cotunneling processes and for higher  $\{|\downarrow0\rangle,|0\uparrow\rangle\}$ and consistently transport channels labeled  by these quantum numbers are active with conductance per channel close to the unitary limit $e^{2}/h$ for SU(2) lines and $1/2 (e^{2}/h)$ per channel in SU(4) point. Figure \ref{fig:fig1}(d)  illustrates partial conductances for the field induced SU(3) Kondo effect. The curve is drawn for the fixed value of the gate voltage. Partial conductances of three channels take the values $3/4 (e^{2}/h)$ each.
\begin{figure}
\includegraphics[width=0.48\linewidth]{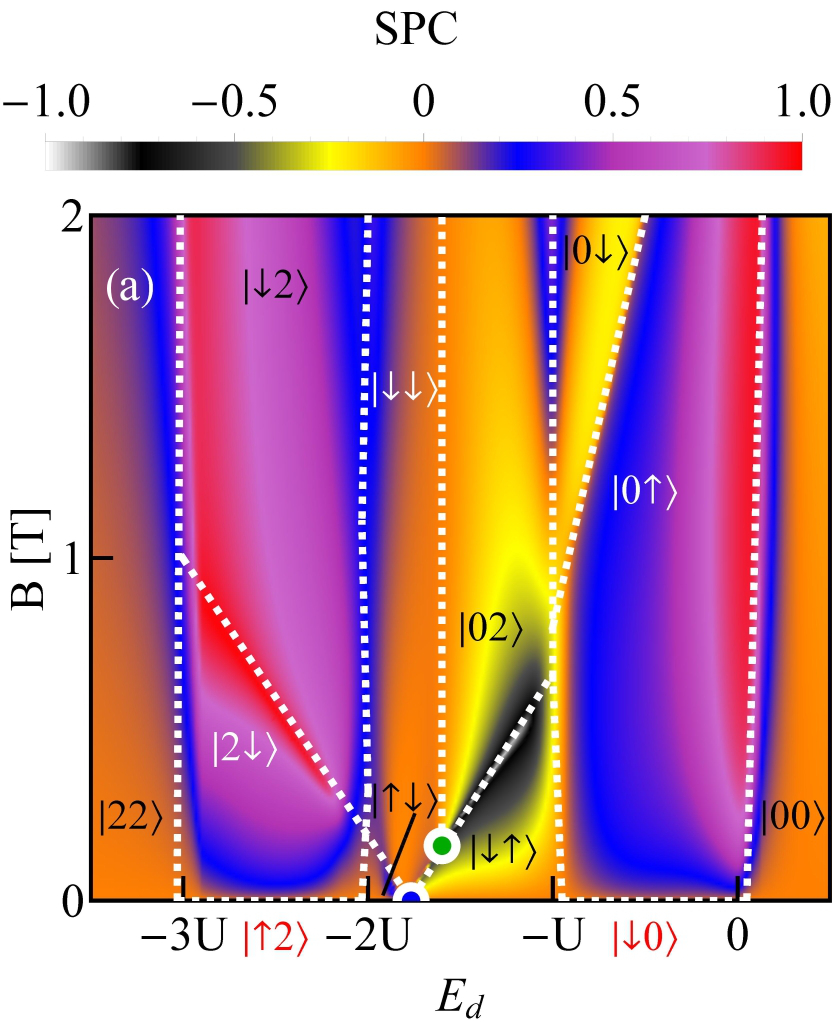}
\includegraphics[width=0.48\linewidth]{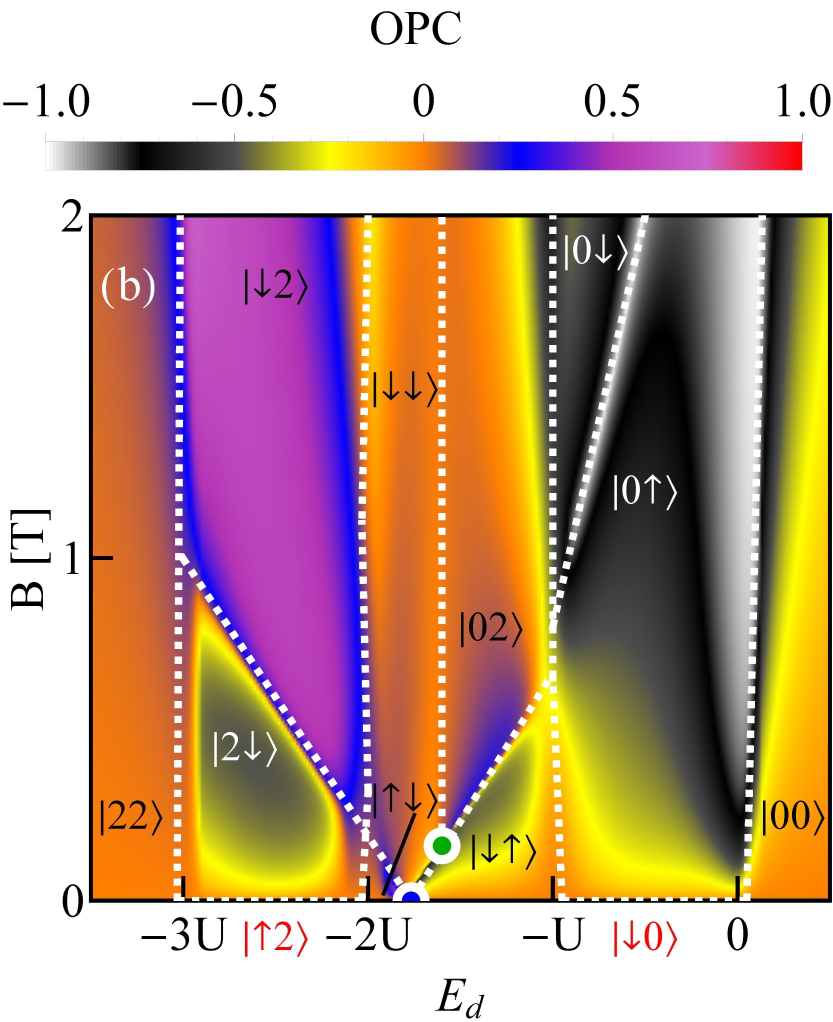}\\
\includegraphics[width=0.48\linewidth]{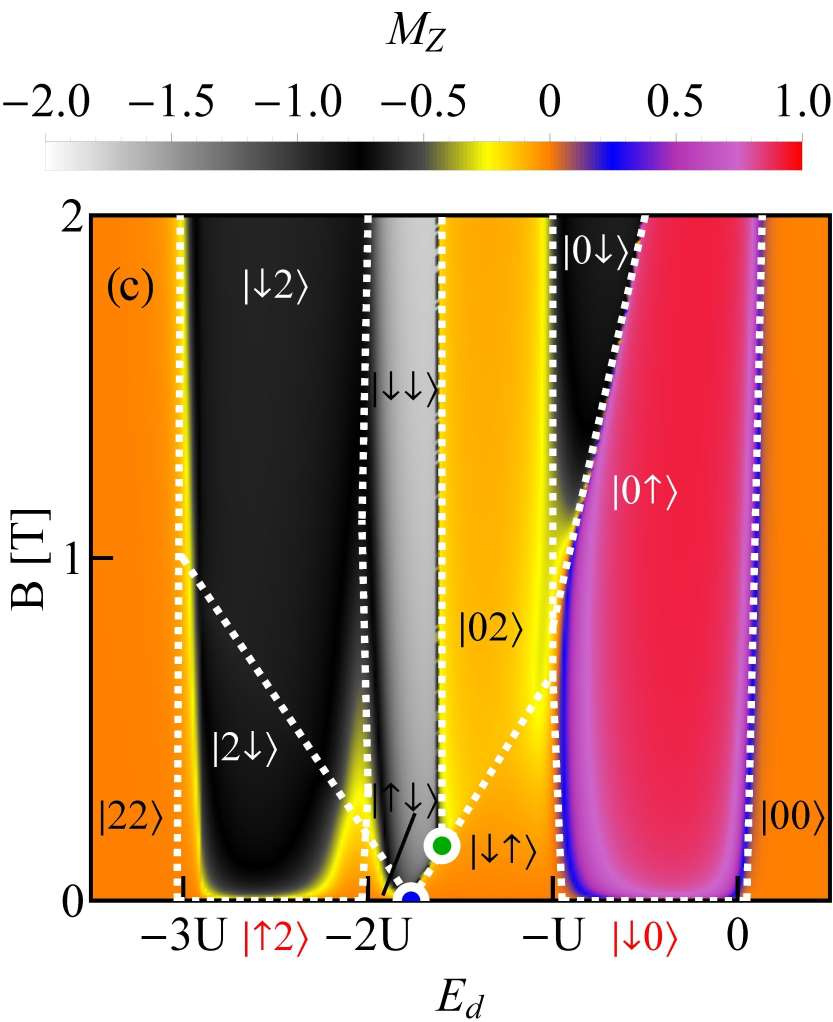}
\includegraphics[width=0.48\linewidth]{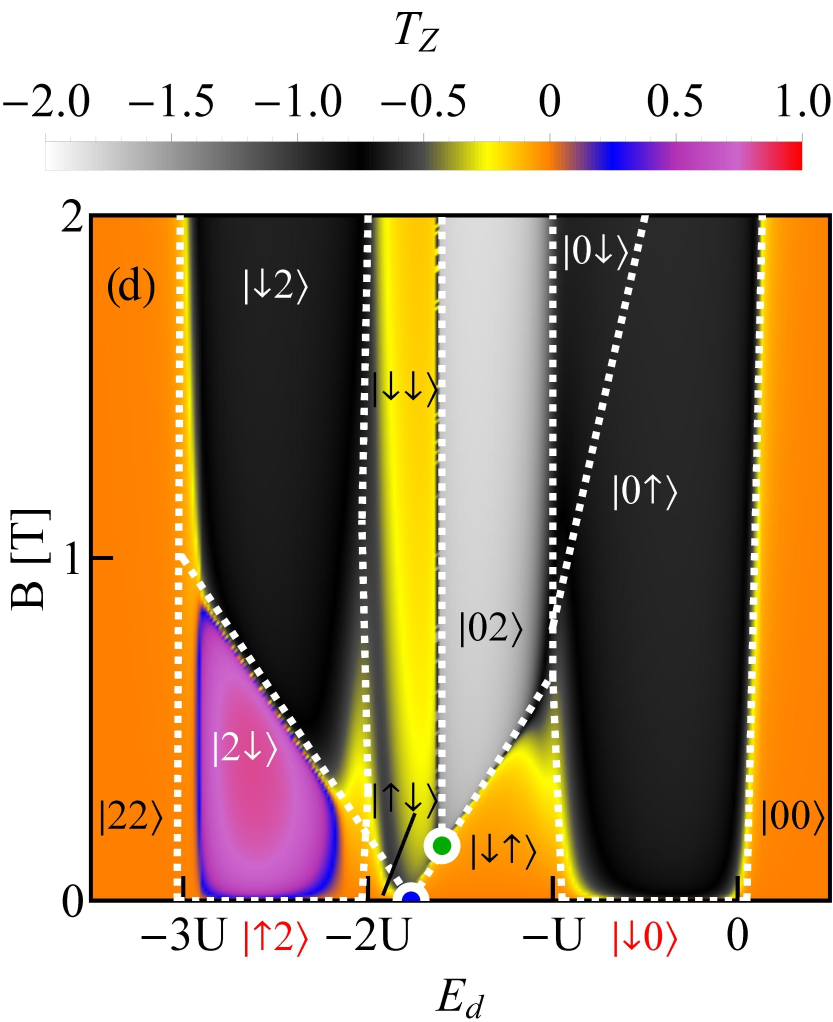}
\caption{\label{fig:fig3} (a),(b) Spin (SPC) and orbital polarization (OPC) maps of CNTQD(15,12). (c),(d) Spin ($M_{Z}$) and orbital  pseudo-spin  ($T_{Z}$) magnetization diagrams of CNTQD(15,12). Dotted white lines represent degeneracy lines.}
\end{figure}
\begin{figure}
\includegraphics[width=0.8\linewidth]{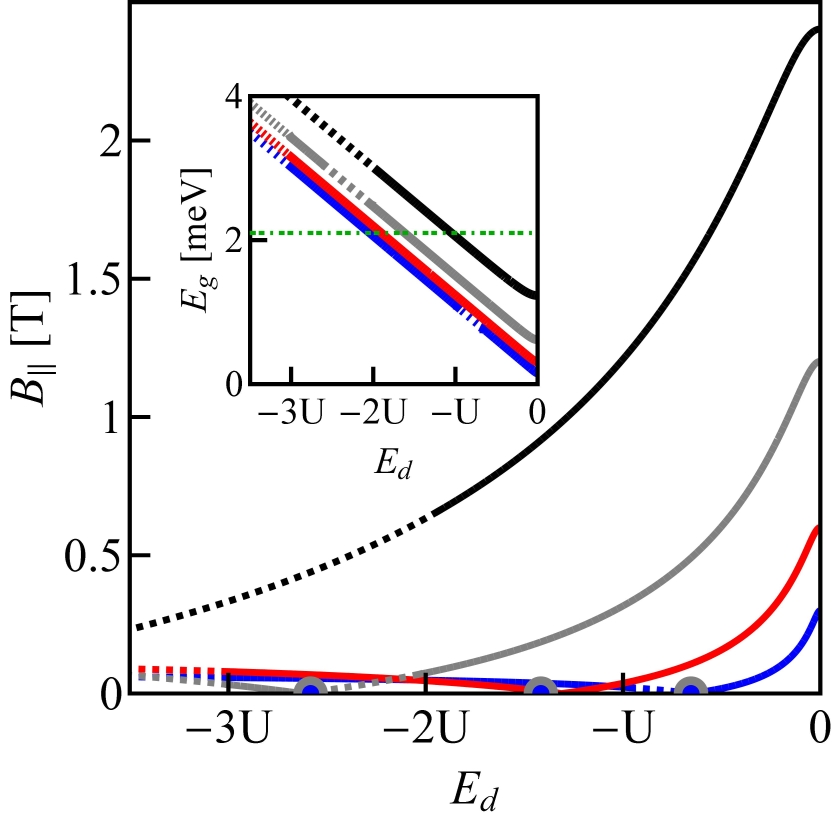}
\caption{\label{fig:fig4} Dependencies of SU(3) Kondo lines of CNTQD(15,12) on magnetic field and on site energy plotted for several SO parameters. Inset presents dependencies of SU(3) lines on site energy and gap. Solid lines correspond to SU(3) Kondo states and dotted parts of the lines indicate the regions where Kondo correlations are destroyed. Green horizontal dotted line in the inset shows the equilibrium value of the gap for the unstrained nanotube. Blue points represent the SU(4) high-symmetry Kondo solution ($B_{\|}=0$). Blue, red, gray and black lines are for $\delta=1/4, 1/2, 1$ and $2$ meV nm ($U=6$ meV).}
\end{figure}
\begin{figure}
\includegraphics[width=0.48\linewidth]{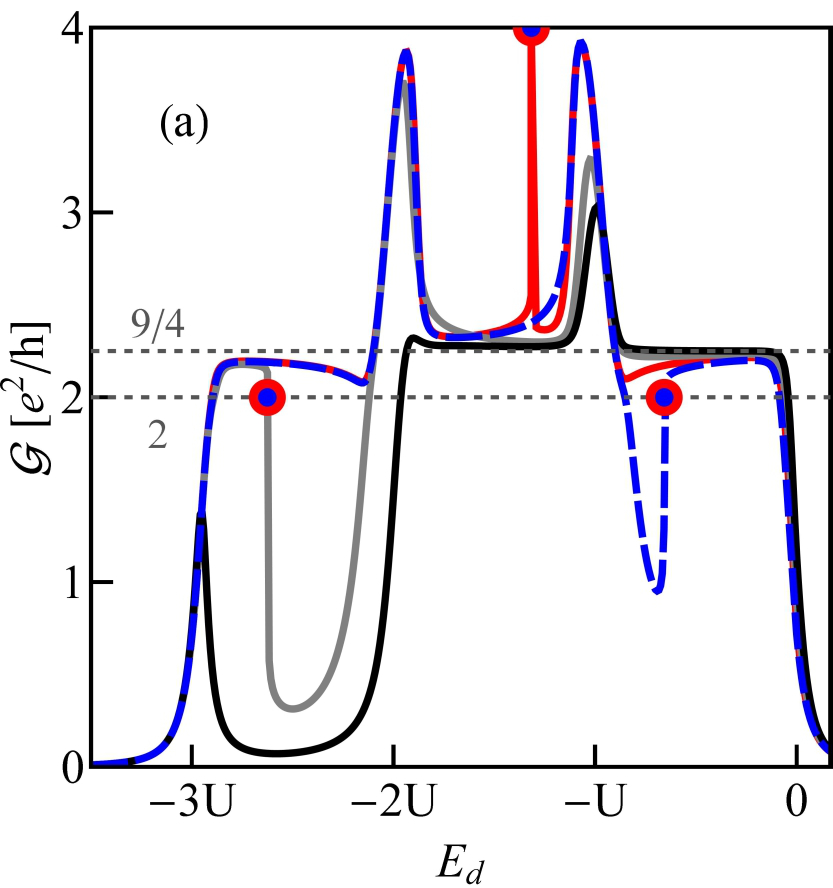}
\includegraphics[width=0.48\linewidth]{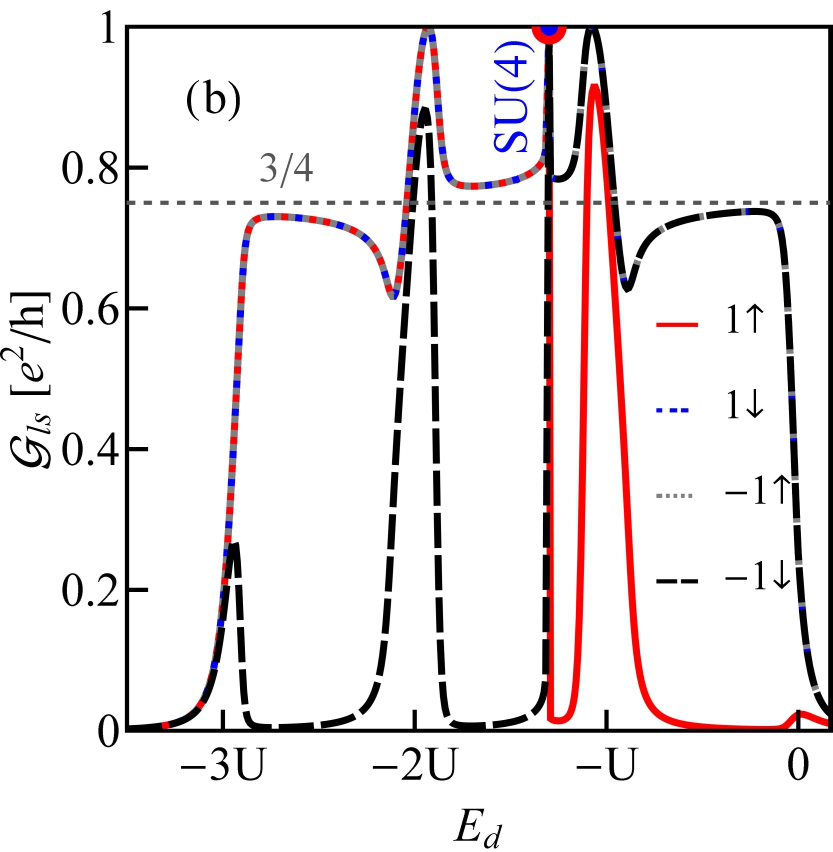}\\
\includegraphics[width=0.48\linewidth]{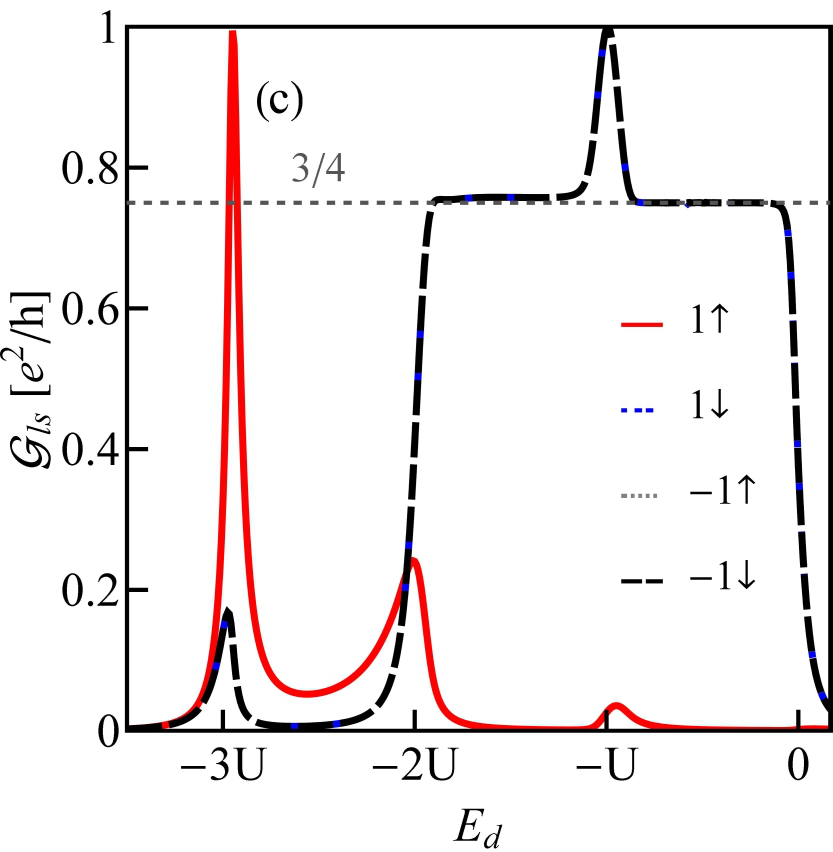}
\includegraphics[width=0.48\linewidth]{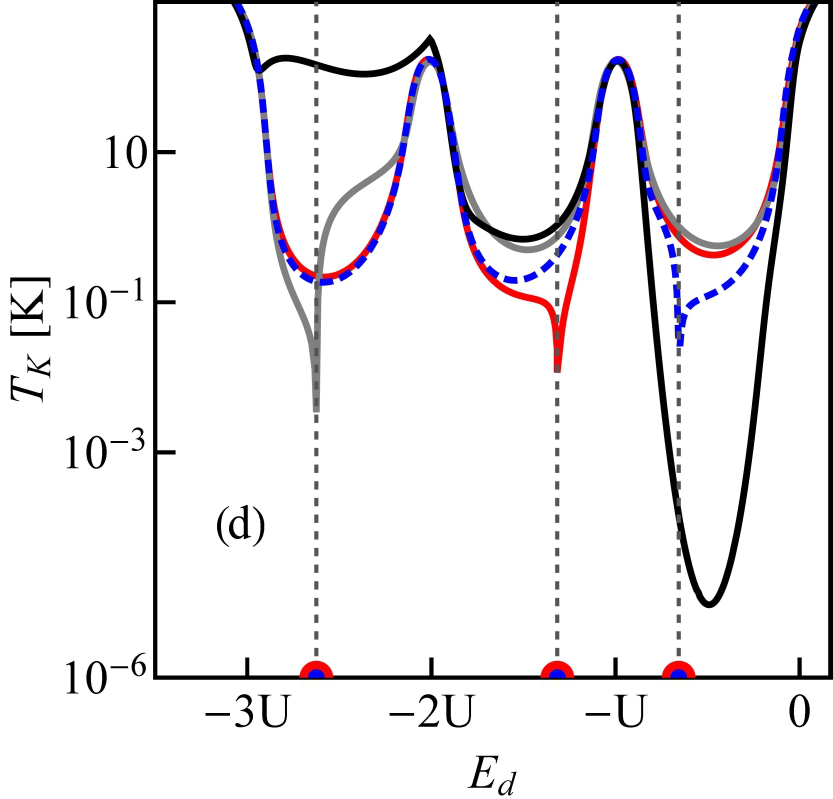}
\caption{\label{fig:fig5} (a),(d) Total conductances (a) and Kondo temperature (d) of CNTQD(15,12) plotted as a function of site energy. Blue, red, gray and black lines are for $\delta = 1/4, 1/2, 1$ and $2$ meV nm respectively. (b) Partial conductances for $\delta = 1/2$ meV nm. (c) Partial conductances for $\delta = 1$ meV nm ($U=6$ meV, $\Gamma= 0.03$ meV).}
\end{figure}
In the example discussed above, the high symmetry points  were located in the region of single occupation, but in different nanotubes they can be located in different occupation areas. For brevity of presentation we show on figure \ref{fig:fig2} only four examples of conductance maps of  CNTQDs  differing in mutual positions of high symmetry points. As it is seen  SU(4) points appear from the intersection of four SU(2) lines and this occurs in zero magnetic field. This is a general condition and  it results from the time inversion symmetry. The appearance  of SU(4) Kondo state in nanotube with finite SO interaction is a surprising result, but it happens due to the gate induced reconstruction of the dot states, which for a certain gate voltage  compensates the changes induced by SO coupling.  SU(3) Kondo effect is field induced and threefold degeneracy point occurs at the intersection of three lines of double degeneracy. To indicate  which quantities fluctuate in the presented Kondo effects, we have also marked on figure \ref{fig:fig2} the corresponding  ground states appearing in  different areas of the maps. The degeneracy lines corresponding to the same occupation are the borders between different ground states in a given Coulomb valley and it is the effective fluctuations of  these degenerated states induced by  cotunneling  processes that lead to the formation of Kondo resonances.

\begin{figure}
\includegraphics[width=0.48\linewidth]{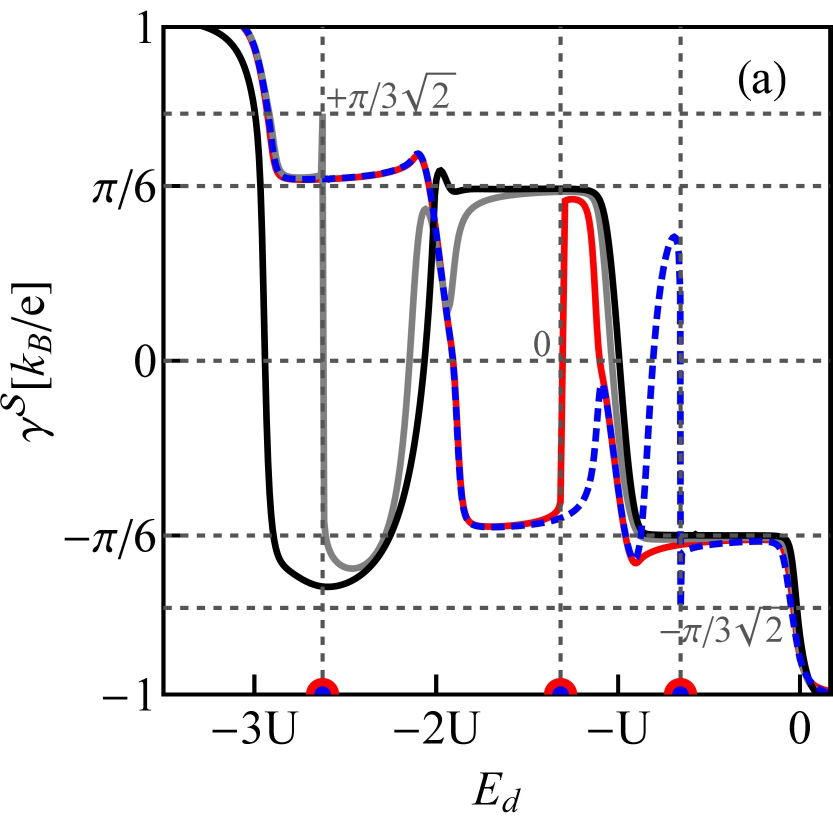}
\includegraphics[width=0.48\linewidth]{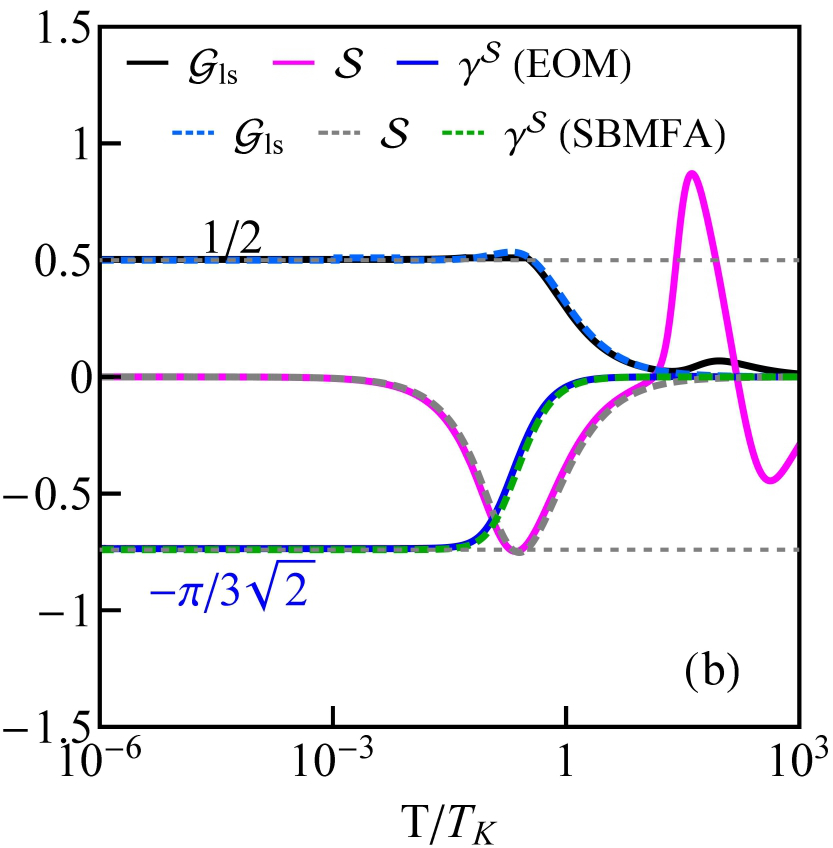}
\caption{\label{fig:fig6} Thermoelectric quantities and conductance of CNTQD(15,12): (a) Linear thermopower coefficient $\gamma^{{\cal{S}}}$ vs. $E_{d}$. Blue, red, gray and black lines are for $\delta = 1/4, 1/2, 1$ and $2$ meV nm respectively. (b) Partial conductance ${\cal{G}}_{ls}$, TEP (${\cal{S}}$), and  $\gamma^{{\cal{S}}}$ as a function of normalized temperature $T/T_{K}$ for $\delta = 1/4$ meV nm. The dotted and solid lines present results of SBMFA and EOM method respectively ($U=6$ meV, $\Gamma=0.03$ meV, $\beta=37$ meV $nm^{2}$).}
\end{figure}
\begin{figure}
\includegraphics[width=0.48\linewidth]{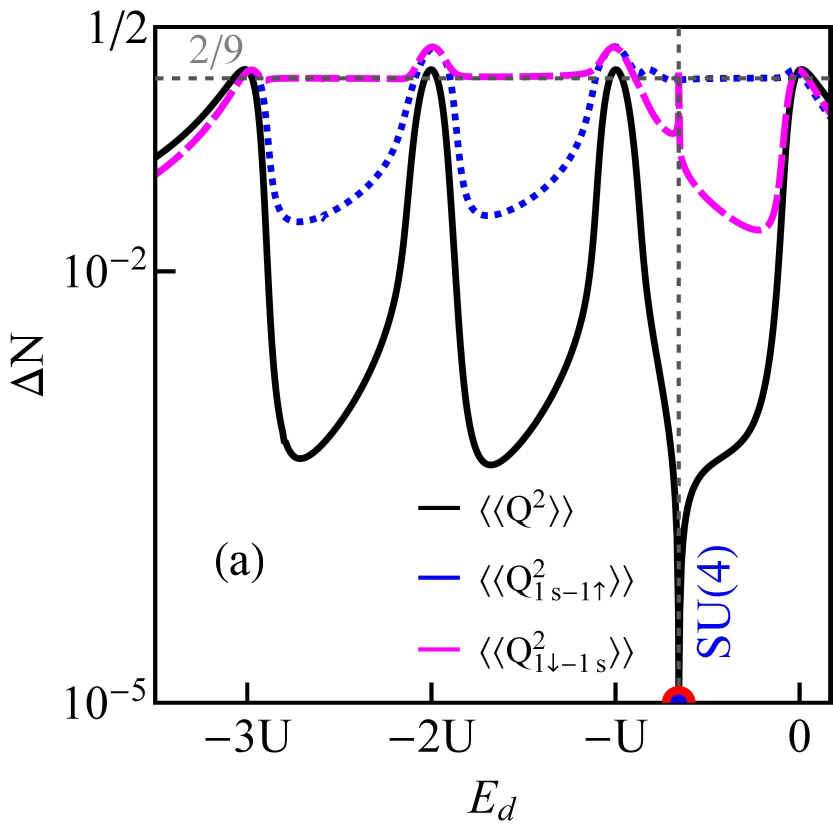}
\includegraphics[width=0.48\linewidth]{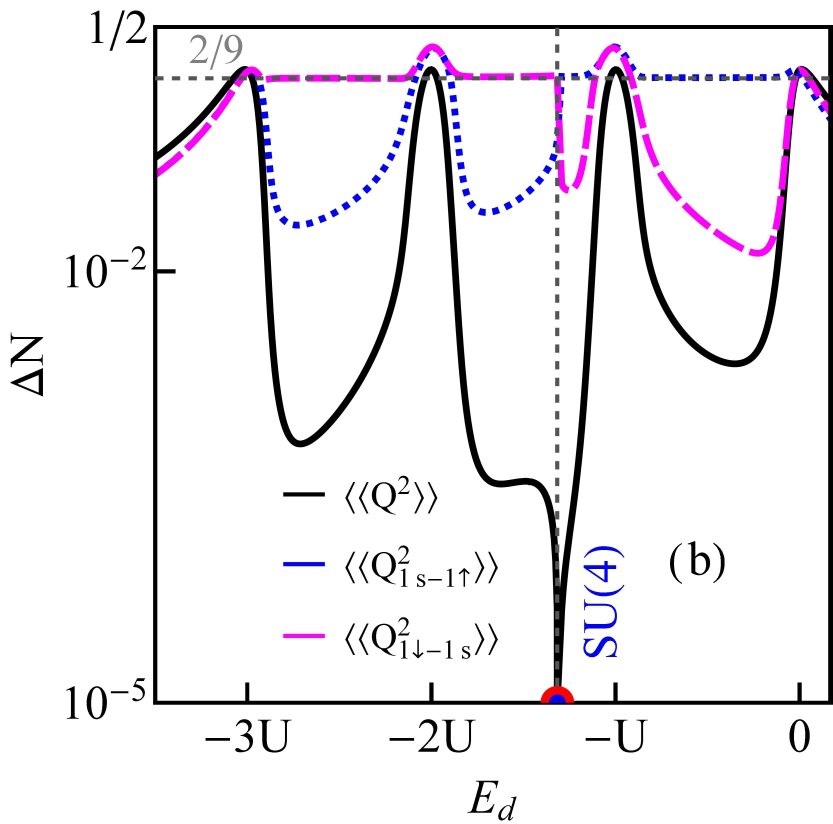}\\
\includegraphics[width=0.48\linewidth]{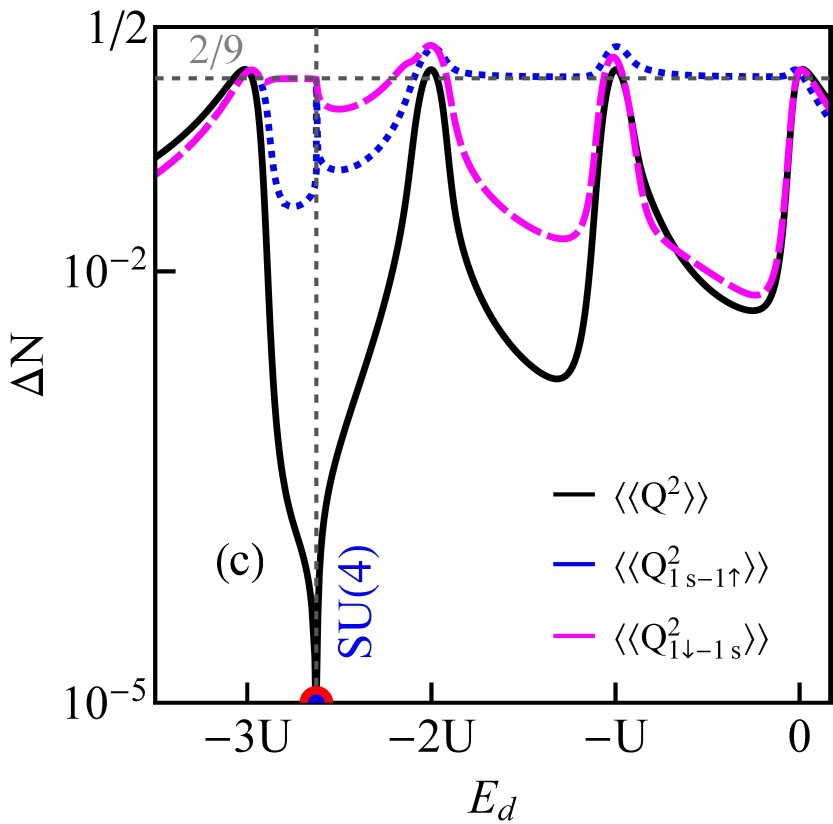}
\includegraphics[width=0.48\linewidth]{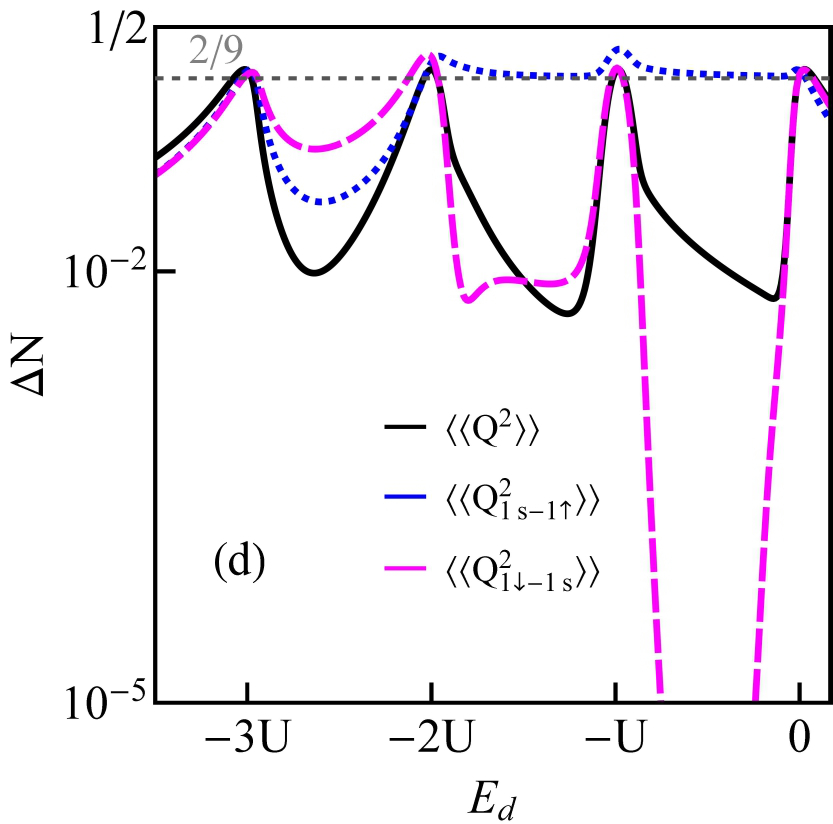}
\caption{\label{fig:fig7} Charge and spin-orbital fluctuations of CNTQD(15,12) plotted for (a) $\delta = 1/4$ meV nm, (b) $\delta = 1/2$ meV nm, (c) $\delta = 1$ meV nm, (d) $\delta = 2$ meV nm ($U=6$ meV, $\Gamma=0.03$ meV, $\beta=37$ meV $nm^{2}$).}
\end{figure}
The first map (figure \ref{fig:fig2}(a)) refers to the carbon  nanotube quantum dot  already discussed CNTQD(24,21), but now the conductance  is shown for a wide range of magnetic fields  and in the entire range of occupation of the first shell. Different gate dependent SU(2) conduction lines reaching values close to $2e^{2}/h$ are seen:  in 1e region spin Kondo effect with fluctuating states  $|0\downarrow\rangle$ and $|0\uparrow\rangle$ exhibiting orbital polarization, valley  Kondo effect with $|0\downarrow\rangle$ and $|\downarrow0\rangle$ states exhibiting spin polarization and spin-valley Kondo effect with fluctuations of  $|\downarrow0\rangle$ and $|0\uparrow\rangle$, where both spin and orbital moments are quenched. SU(3)  Kondo effect is caused by cotunneling induced fluctuations of $|0\downarrow\rangle$, $|0\uparrow\rangle$ and $|\downarrow0\rangle$  states and the resulting resonance is spin and orbital polarized. SU(4) Kondo screening results from effective fluctuations of all four spin-orbital states $|ls\rangle$.
\begin{figure}
\includegraphics[width=0.48\linewidth]{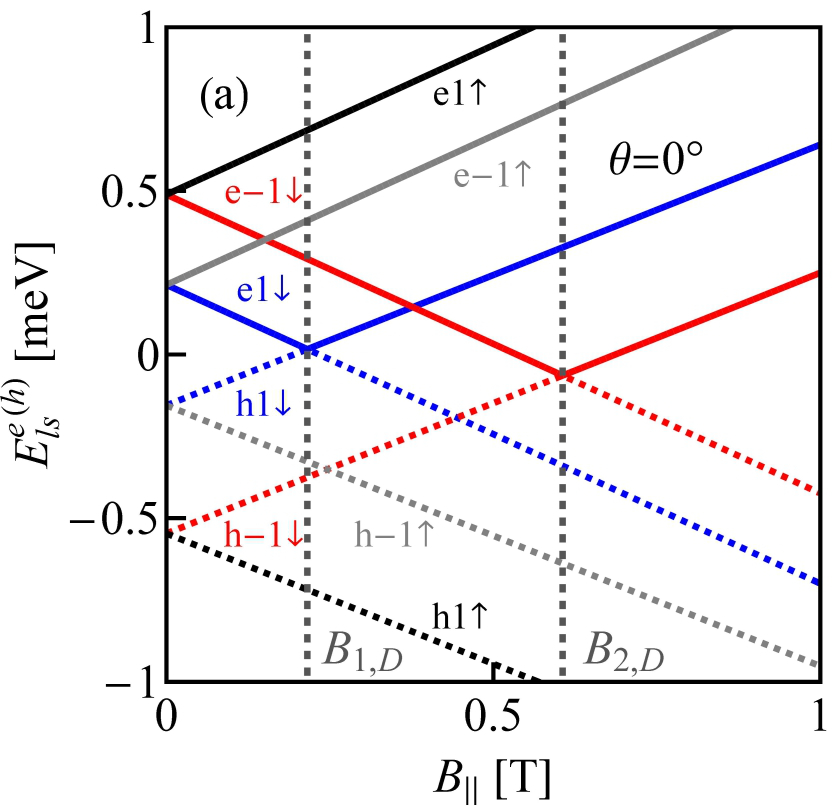}
\includegraphics[width=0.48\linewidth]{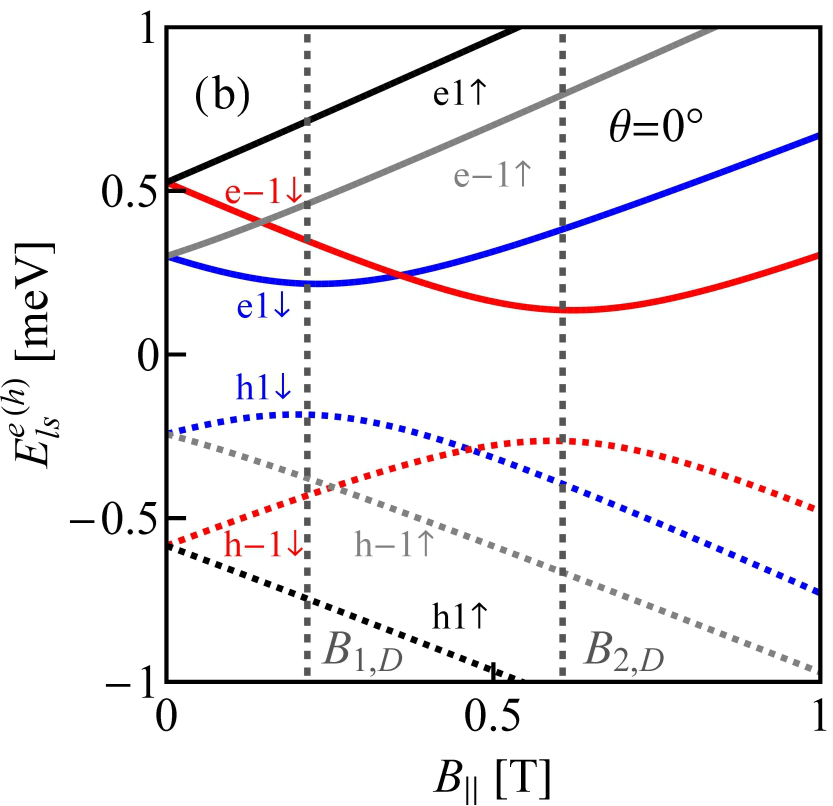}\\
\includegraphics[width=0.48\linewidth]{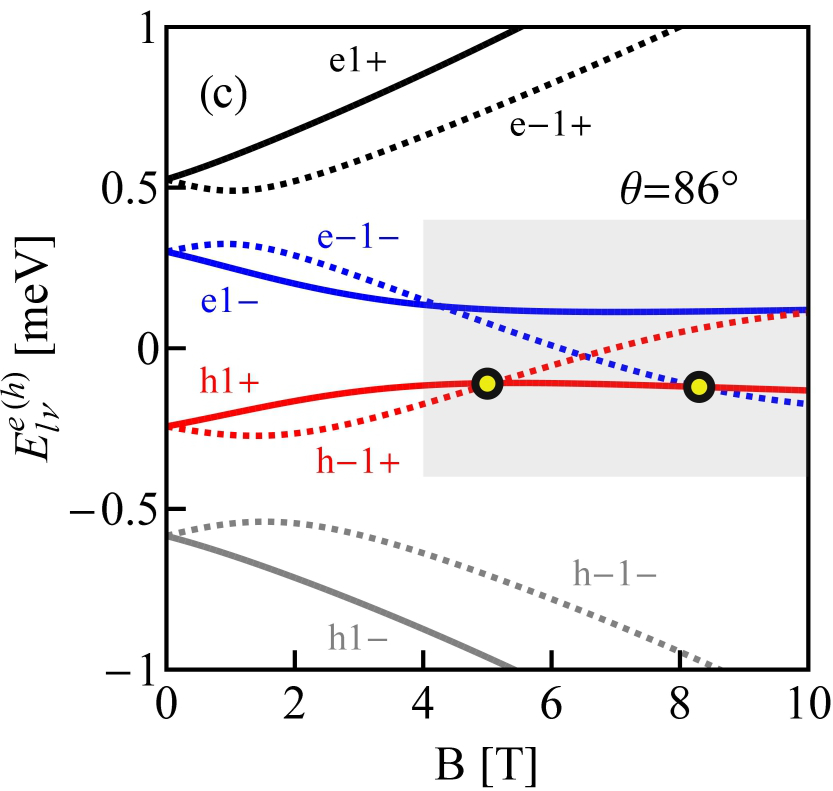}
\includegraphics[width=0.48\linewidth]{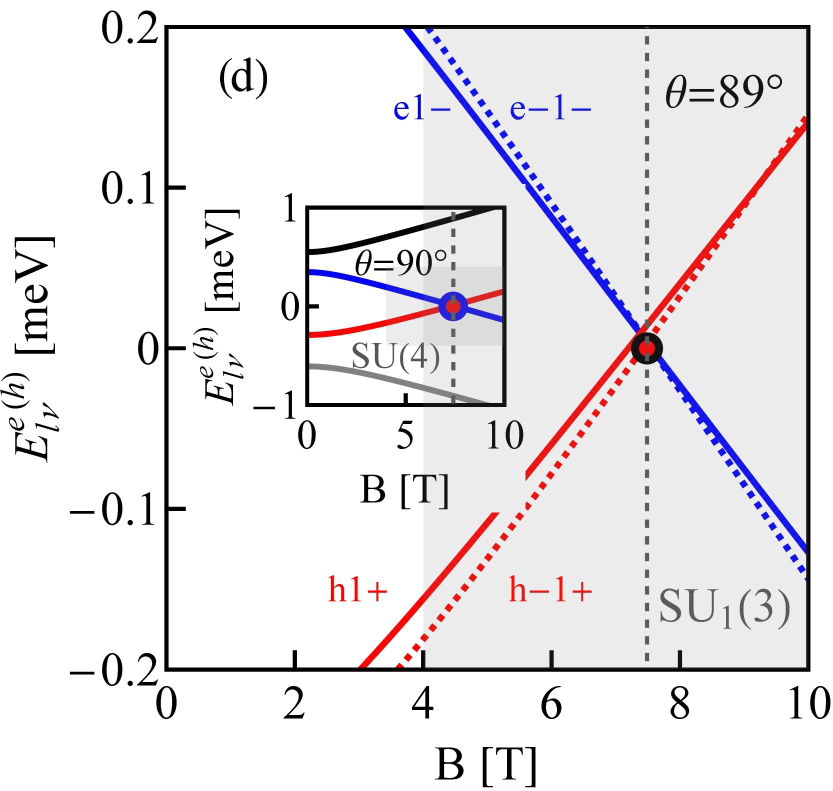}
\caption{\label{fig:fig8} Electron and hole states of (a) carbon nanotube C(33,30) and of quantum dot CNTQD(33,30) (b,c,d), (b) parallel magnetic field ($\theta = 0^{\circ}$) and in slanting magnetic fields:  (c) $\theta = 86^{\circ}$, (d) $\theta = 89^{\circ}$ ($\delta=3/2$ meV nm, $\beta=37$ meV $nm^{2}$, $E_{d} = -0.2$ meV).}
\end{figure}
\begin{figure}
\includegraphics[width=0.48\linewidth]{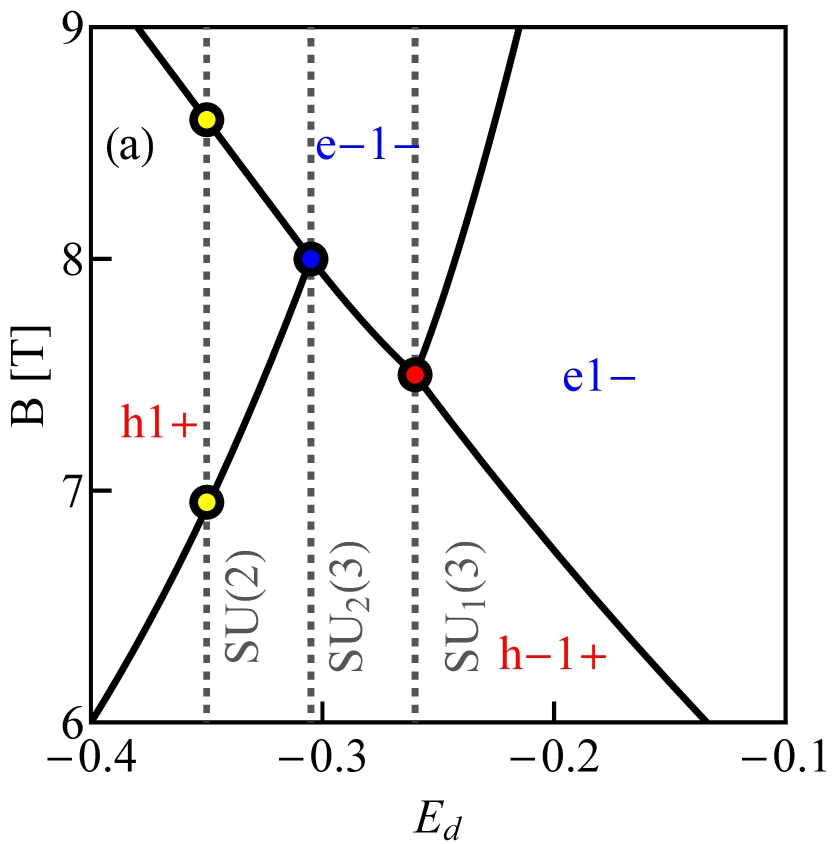}
\includegraphics[width=0.48\linewidth]{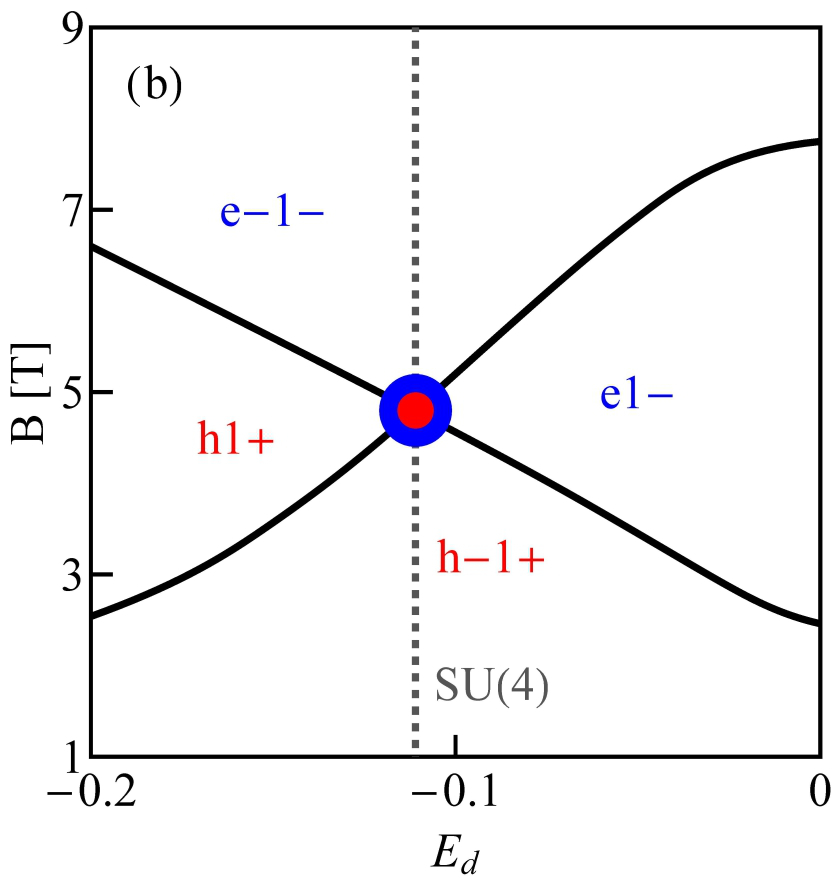}\\
\includegraphics[width=0.48\linewidth]{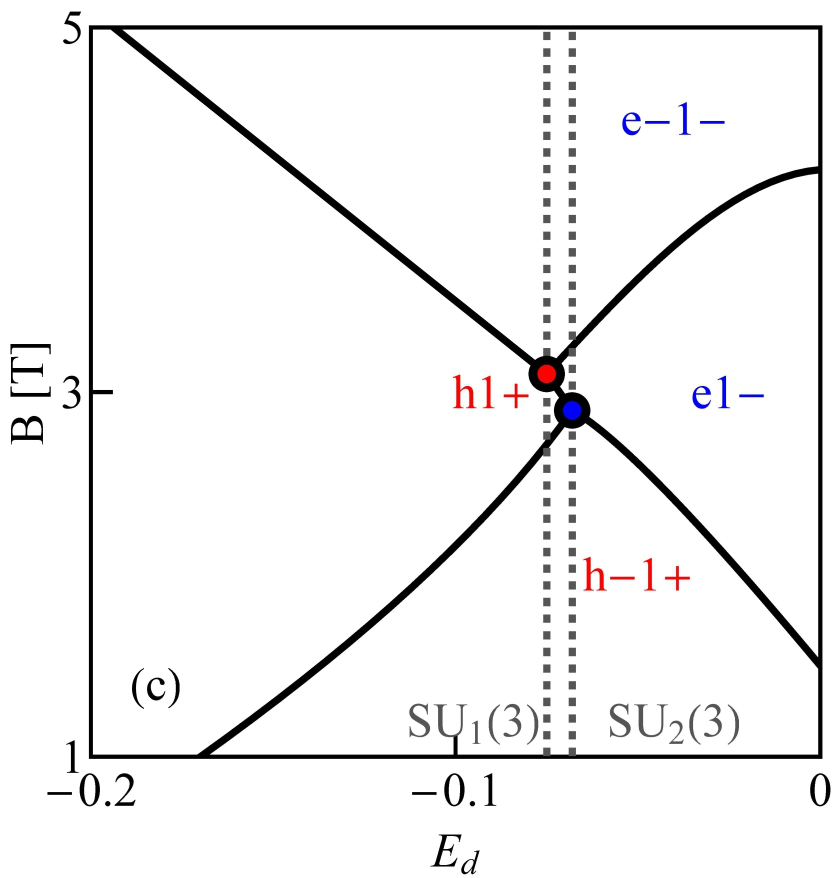}
\includegraphics[width=0.48\linewidth]{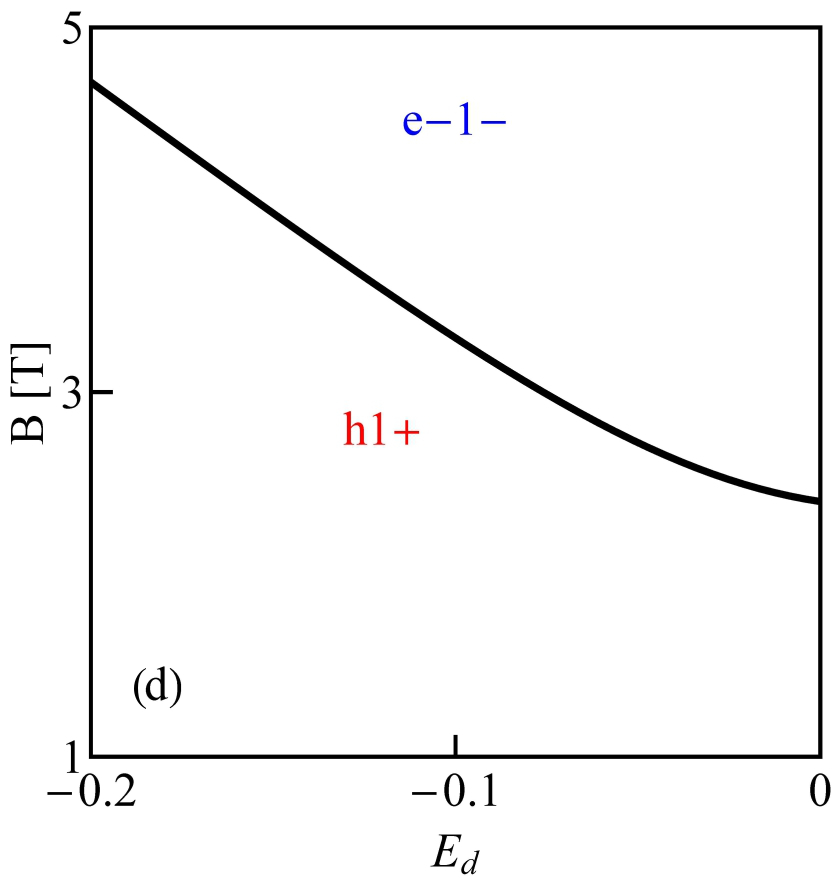}
\caption{\label{fig:fig9} Electron-hole ground state diagrams of CNTQD(33,30) in slanting magnetic field ($\theta = 89^{\circ}$) for different SO parameters : (a) $\delta=3/2$ meV nm, (b) $\delta=0.9$ meV nm, (c) $\delta=0.8$ meV nm and (d) $\delta=0.4$ meV nm ($U = 6$ meV, $\beta=37$ meV $nm^{2}$).}
\end{figure}
In 2e valley we observe two-electron spin polarized  spin Kondo effect with  $|\uparrow\downarrow\rangle$  and $|\downarrow\downarrow\rangle$ dot states engaged and  spin-orbit effect in 3e valley with fluctuating  $|\downarrow2\rangle$ and $|2\downarrow\rangle$  states and screened spin and orbital moments.
On the next map representing conductance of quantum dot formed in C(39,24)  tube SU(4) point lies in 1e valley and SU(3) in the region of double occupancy (figure \ref{fig:fig2}(b)). On figure \ref{fig:fig2}(c), in turn, which presents conductance of CNTQD(15,12)  both SU(4) and SU(3) points are located in 2e valley and the last  map (figure \ref{fig:fig2}(d))  presenting conductance of CNTQD(48,18)  has SU(4) point in 1e valley and SU(3) point in 3e valley. As already mentioned, the examples presented above do not exhaust all possible locations of high symmetry points. Figure \ref{fig:fig3}(a) presents example of spin polarization of conductance  corresponding to conduction map  figure \ref{fig:fig1}(c). Interesting feature from spintronic  point of view is the occurrence of Kondo lines with high positive and negative conductance polarizations between which one can switch by gate voltage (spin filter).

For any narrow gap nanotube  one can move the SU(3) point between different occupation  areas by changing magnetic field and the value of the gap, the latter change  can be induced by strain. The examples of SU(3) lines drawn for  CNTQD(15,12)  for several assumed SO parameters are presented on figure \ref{fig:fig4}. The required magnetic field for the occurrence of threefold degeneracy for  given gate voltage can be read from the main picture, while the gap can be read from the inset. The solid lines present SU(3) Kondo solutions and the dotted parts of the lines correspond to the situation, where Kondo correlations are destroyed.  The fourfold degeneration points are also marked and they appear when the SU(3)  line touches the $B = 0$ line. The gate dependence of corresponding conductances are shown on figure \ref{fig:fig5}(a). Horizontal dashed lines indicate the characteristic  limit of SU(3) Kondo conductance ${\cal{G}} = 9/4 (e^{2}/h)$ and unitary value for SU(4) Kondo effect in odd valleys ${\cal{G}} = 2 (e^{2}/h)$. The centers of Kondo resonances are shifted from $E_{F}$ in odd valleys, shifted towards lower gate voltages in 1e valley and towards higher voltages in 3e valley. In 2e valley conductance of SU(4) Kondo point reaches value ${\cal{G}} = 4 (e^{2}/h)$  due to the presence of six degenerate states. Kondo state is formed due to cotunneling induced fluctuations between all these states. SU(4) Kondo resonance in this case is centered at $E_{F}$. For gate voltage intervals where Kondo correlations are destroyed  the drop of conductance is observed. We also present examples of  partial conductances for two values of SO coupling:  $\delta = 2$ meV nm, for which SU(4) point is pushed out of the first shell and for $\delta = 0.5$ meV nm, for which  SU(4) point locates in 2e valley. In the former case we observe partial SU(3) conductance  reaching almost value $3/4 (e^{2}/h)$, identical for the three spin channels and in the latter case SU(4) point divides the Kondo SU(3) line in 2e region into two parts associated with two different resonances of the same symmetry. One SU(3) resonance associated with fluctuations of the  states $\{|\downarrow\uparrow\rangle, |\downarrow\downarrow\rangle, |02\rangle\}$ and other with fluctuations of $\{|\uparrow\downarrow\rangle, |\downarrow\downarrow\rangle, |02\rangle\}$. The quantity that clearly  reflects the symmetries of the  many-body resonances and their electron or hole character within the shell is a linear thermoelectric coefficient of thermopower defined as $\gamma^{{\cal{S}}}=lim_{T,V_{sd}\rightarrow0}\frac{{\cal{S}}T_{K}}{2\pi T}\approx\frac{-k_{B}\pi}{3e}\frac{\widetilde{E}_{ls}}{T_{K}}$ \cite{Krychowski1}, where Kondo temperature $T_{K}$ is given by the center and the width of Kondo resonance $T_{K}
=\sqrt{\widetilde{E}^{2}_{ls}+\widetilde{\Gamma}^{2}_{ls}}$, with $\widetilde{E}_{ls}=E^{e}_{ls}+\lambda_{ls}$ and $\widetilde{\Gamma}_{ls}=\Gamma z^{2}_{ls}$ \cite{Kotliar}, where $ls$ label the dot states active in Kondo processes. The plots of $\gamma^{{\cal{S}}}$ are presented on figure \ref{fig:fig6}(a). The dashed red horizontal lines correspond to the characteristic limits for a given symmetry $\pm\pi/6(k_{B}/e)$  for SU(3) Kondo effect and $\pm\pi/3\sqrt{2}(k_{B}/e)$ for SU(4) in odd valleys. In 1e valley, regardless of symmetry, $\gamma^{{\cal{S}}}$ is always negative (electron nature), and in  3e valley  positive (hole character).  In 2e valley $\gamma^{{\cal{S}}} = 0$ for SU(4) Kondo state and for SU(3) symmetry the change of the sign of $\gamma^{{\cal{S}}}$ is observed, which reflects transition between the previously mentioned two SU(3) Kondo states. Switching between different Kondo SU(3) states is also visible in the spin-orbital fluctuations associated with the given resonances. Figure \ref{fig:fig7} presents charge fluctuations and spin-orbital fluctuations corresponding to two types of SU(3) Kondo effects occuring in the system. The corresponding second cumulants are defined as follows:
\begin{eqnarray}
&&\langle\langle Q^{2}\rangle\rangle=\langle(N-\langle N\rangle)^{2}\rangle = \langle N^{2}\rangle-\langle N\rangle^{2}\nonumber\\
&&\langle\langle Q_{1,-1\uparrow}^{2}\rangle\rangle = \langle\sum_{s}(N_{1s}+N_{-1\uparrow})^{2}\rangle-\nonumber\\
&&\langle \sum_{s}(N_{1s}+N_{-1\uparrow}) \rangle^{2}\\
&&\langle\langle Q_{1\downarrow,-1}^{2}\rangle\rangle = \langle\sum_{s}(N_{1\downarrow}+N_{-1s})^{2}\rangle-\nonumber\\
&&\langle \sum_{s}(N_{1\downarrow}+N_{-1s}) \rangle^{2}\nonumber
\end{eqnarray}
The above fluctuations can be easily expressed by slave boson operators (see Appendix).
Charge fluctuations and these spin-orbit fluctuations which relate only to the dot  states active in Kondo processes, they have small values and are characterized by clear minima  in the regions  of occurrence of SU(3) resonance. Interesting observation is that  the gate dependence of  Kondo temperature  qualitatively resembles the dependencies of these fluctuations (figure \ref{fig:fig7}, figure \ref{fig:fig5}(d)). Other spin-orbital fluctuations, not related solely to active states in Kondo processes weakly depend on gate voltage and the value they take depends on symmetry.
\begin{figure}
\includegraphics[width=0.48\linewidth]{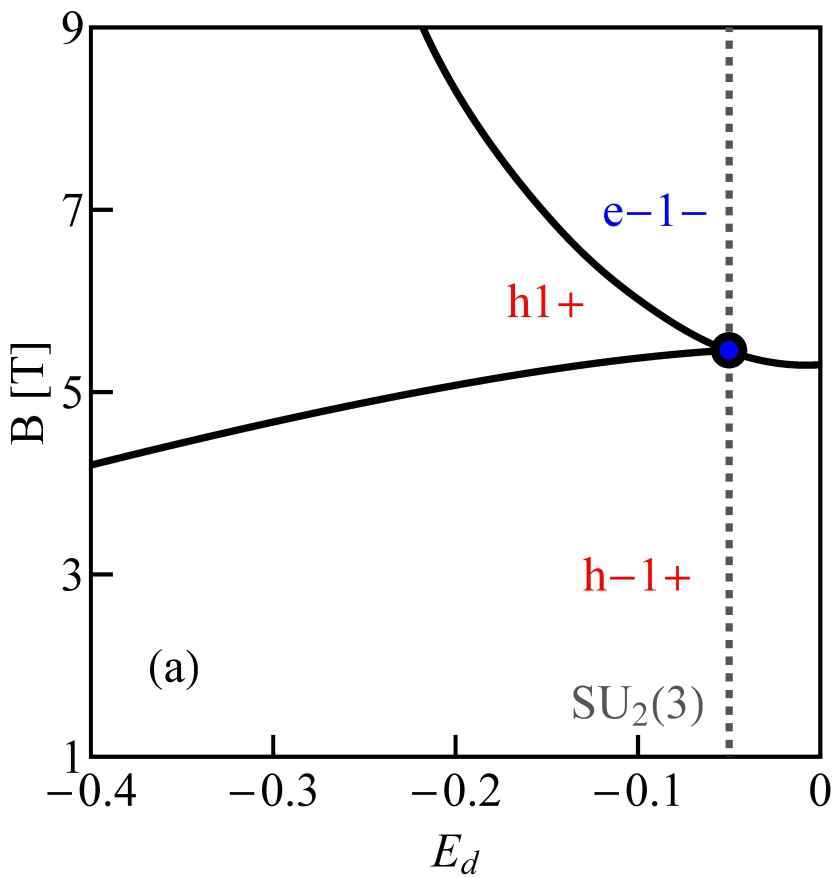}
\includegraphics[width=0.48\linewidth]{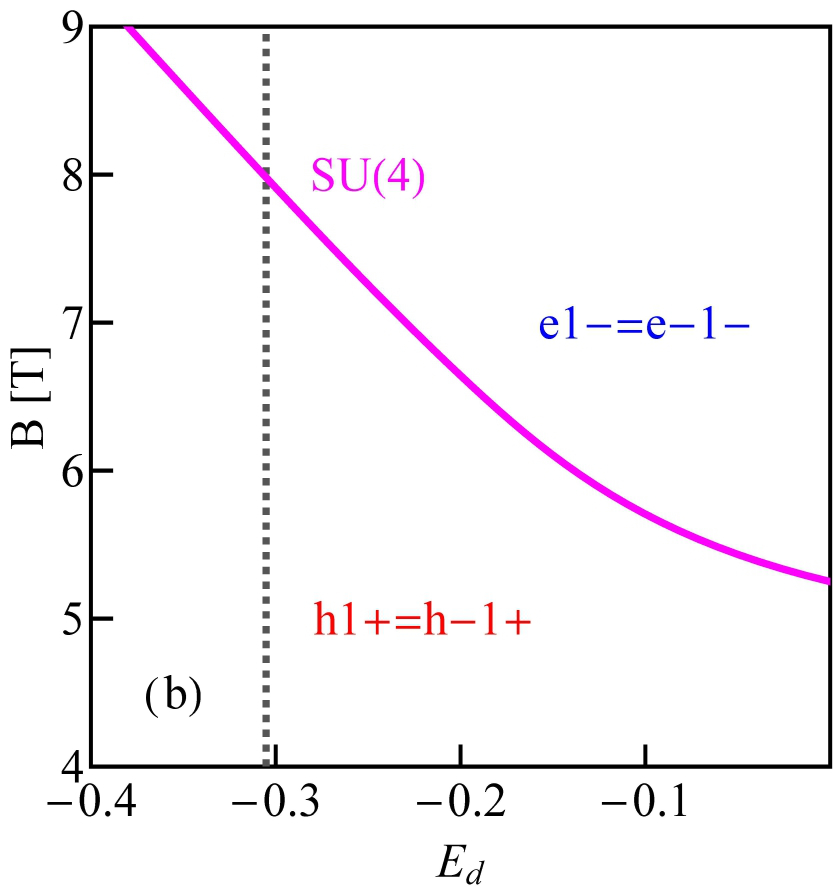}
\caption{\label{fig:fig10} Electron-hole ground state diagrams of  CNTQD(33,30) for (a) $\theta=86^{\circ}$ and (b) $\theta=90^{\circ}$ ($U = 6$ meV, $\delta=3/2$ meV nm, $\beta=37$ meV $nm^{2}$).}
\end{figure}
\begin{figure}
\includegraphics[width=0.48\linewidth]{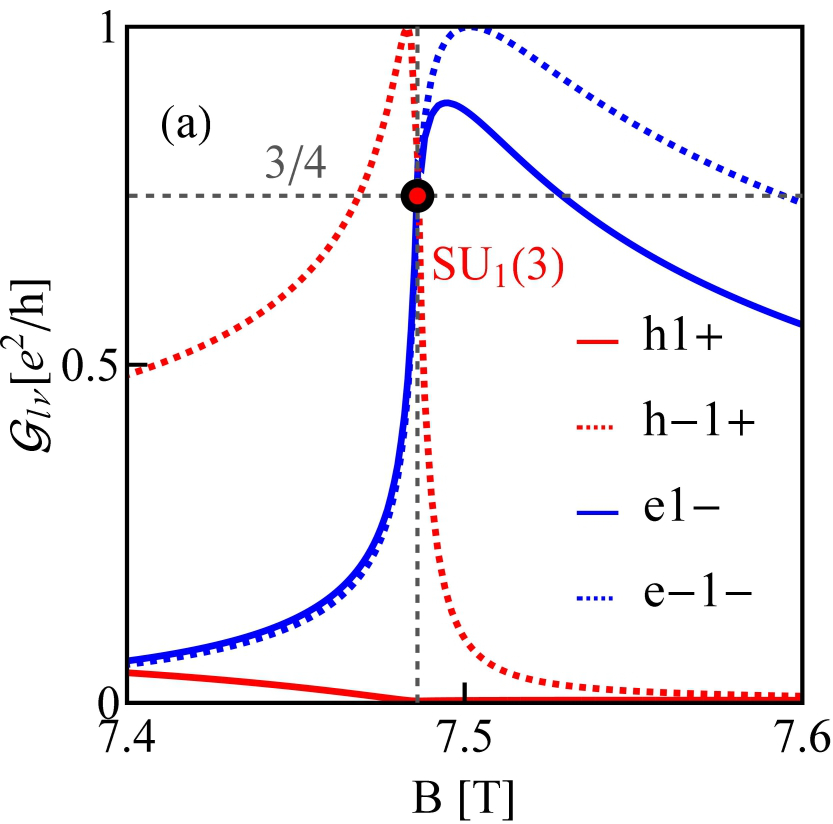}
\includegraphics[width=0.48\linewidth]{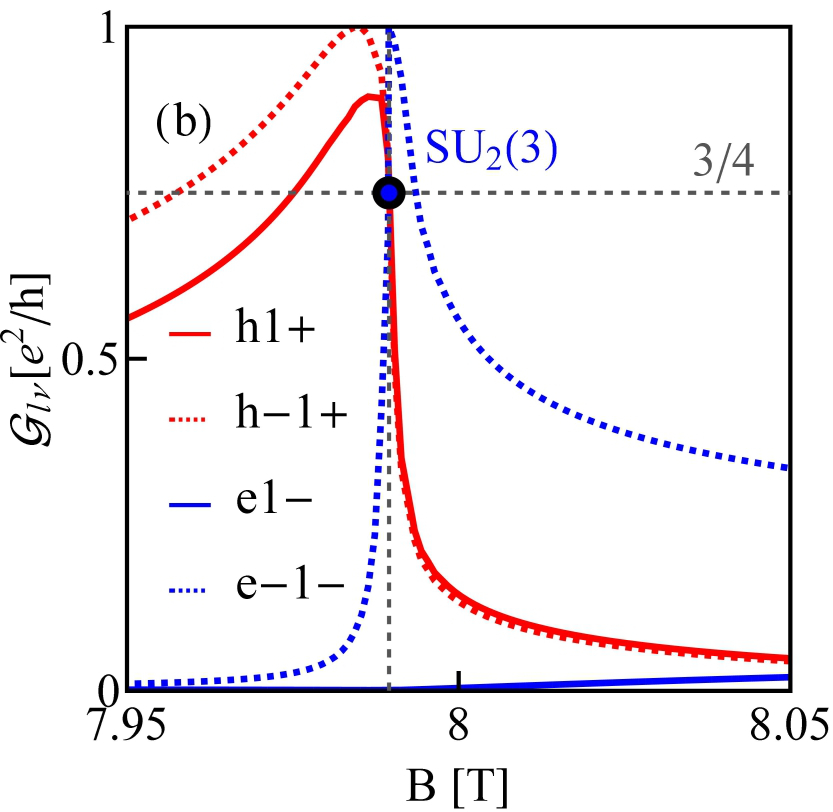}\\
\includegraphics[width=0.48\linewidth]{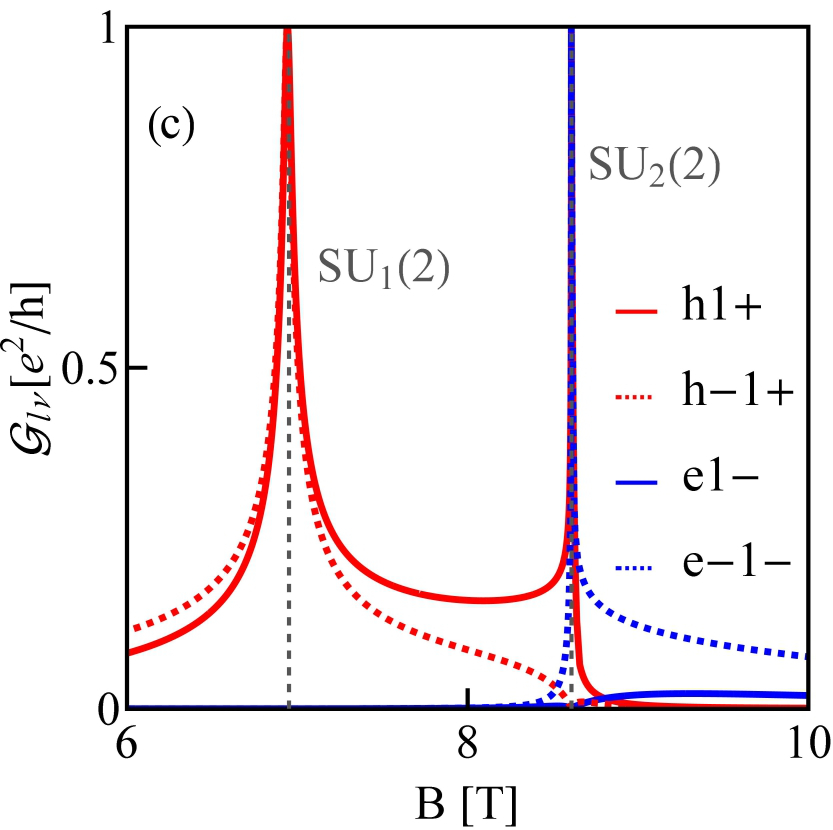}
\includegraphics[width=0.48\linewidth]{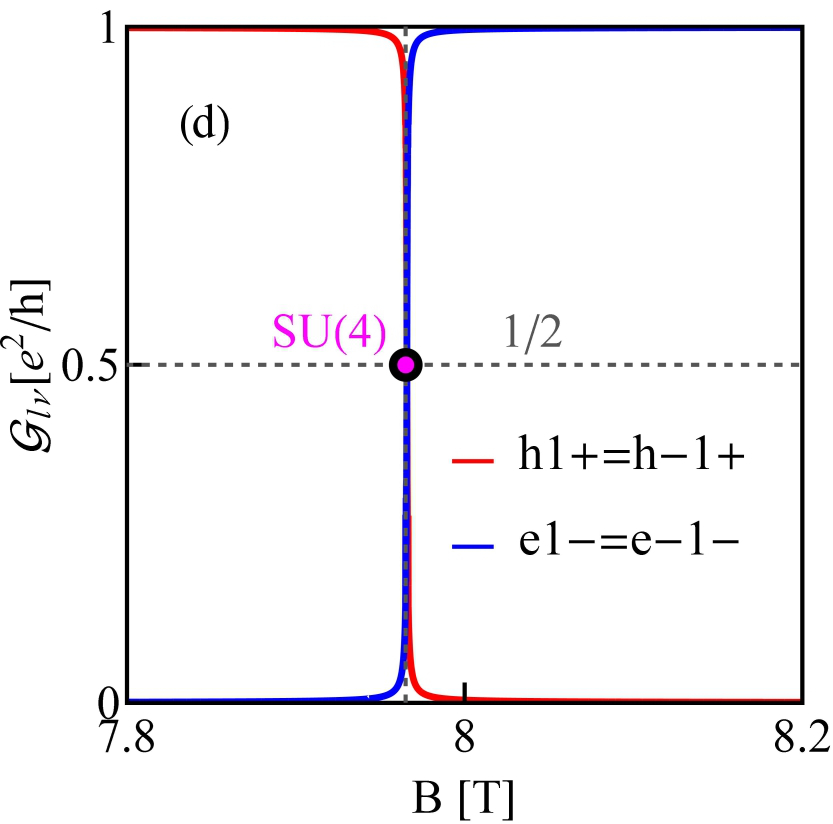}
\caption{\label{fig:fig11} Partial conductances of CNTQD(33,30) in the slanting magnetic fields for  (a,b,c)  $\theta = 89^{\circ}$, curves plotted for cross sections from figure \ref{fig:fig9}(a) designated by gray dotted lines (d) $\theta = 90^{\circ}$ for cross-section marked by dotted lines on figure \ref{fig:fig9}(a) ($\delta = 3/2$ meV nm, $\beta = 37$ meV $nm^{2}$, $\Gamma = 0.01$ meV and $U = 6$ meV).}
\end{figure}
Figure \ref{fig:fig6}(b) shows the example of  temperature dependencies  of conduction, thermoelectric power and $\gamma^{{\cal{S}}}$ coefficient drawn for SU(4) point in 1e valley for CNTQD(15,12) and  calculated by SBMFA  and additionally by  equation of motion method (EOM) with  Lacroix approximation \cite{Lacroix}. The latter approach is introduced to account for high temperature behavior. At low temperatures  both methods reproduce limits characteristic for SU(4) resonance. For $T\ll T_{K}$ thermopower (TEP) approaches a local minimum and for higher temperatures increases and changes sign. This signals the disappearance of Kondo correlations. For still higher temperature another minimum of  TEP is observed, which is due to Coulomb resonance.

In nearly metallic carbon nanotubes magnetic field of several Tesla  closes the gap. An example of field dependencies of lowest electron and highest hole states of nanotube C(33,30)  is presented on figure  \ref{fig:fig8}(a) . The fields at which the gap closes are called Dirac fields.  In CNTQDs parallel magnetic field does not close the gap due to finite confinement energy.   Figure \ref{fig:fig8}(b)    shows the  field dependences of  four  states from the lowest electron shell and four states from the highest hole shell of  the quantum dot CNTQD(33,30). No crossing of electron with hole line is observed for any value of parallel magnetic field ($\theta = 0$). It is worth to observe that the minima or maxima appearing  in the field dependencies of electron  or hole states of quantum dots formed in a given nearly metallic nanotube  occur for the fields  equal to Dirac fields of corresponding infinite nanotube. Crossing of electron and hole energy lines  is observed in slanting fields (figures \ref{fig:fig8}(c), (d)).   For non-parallel fields  the spin states $|\uparrow\rangle$ and $\langle\downarrow|$  are mixed up by perturbation ${\cal{H}}_{\perp} = (1/2)g\mu_{B}B_{\perp}(|\uparrow\rangle\langle\downarrow| + |\downarrow\rangle\langle\uparrow|)$, where $B_{\perp} = Bsin(\theta)$. We will denote the new spin states by  $|+\rangle$, $|-\rangle$. The states are also labeled by orbital index $l$ and we additionally  introduce in the designation of states the letters $e$, $h$ to distinguish between electron and hole states. In this notation the single particle dot states of interest are $|e1-\rangle$, $|e-1-\rangle$, $|h1+\rangle$ and $|h-1+\rangle$.

Although we discuss in the following  many-body resonances only  for a specific example of  CNTQD(33,30), the analysis and conclusions presented below  apply to all the dots formed in quasi-metallic nanotubes. Figures \ref{fig:fig8}(c), (d) show the examples of  field dependencies of electron and hole energies  for $\theta = 86^{\circ}$, $\theta = 89^{\circ}$ and in the inset of figure \ref{fig:fig8}(d) additionally for $90^{\circ}$.  For $\theta = 86^{\circ}$ two ground state double degeneracy points are observed, for lower field hole-hole degeneracy point ($|h1+\rangle,|h-1+\rangle$) and for higher field electron-hole (e-h) degeneracy ($|h1+\rangle,|e-1-\rangle$). Figure \ref{fig:fig8}(d) shows the case of triple degeneracy {$|e1-\rangle$, $|e-1-\rangle$, $|h-1+\rangle$} and the inset of figure \ref{fig:fig8}(d) illustrates  fourfold electron-hole degeneracy occurring in transverse field ($|h1+\rangle$,$|h-1+\rangle$,$|e1-\rangle$,$|e-1-\rangle$). Before discussion of correlation effects let us show how the ground state diagrams of isolated dot change  with the strength of SO interaction or with the orientation  of magnetic field. We restrict to the range of single occupation.  Figure \ref{fig:fig9} presents ground state diagrams for $\theta = 89^{\circ}$ and several values of SO coupling parameter. For $\delta = 3/2$ meV nm apart from four double degeneracy lines SU(2), also two SU(3) points are seen: $SU_{1}(3)$, where the two hole states degenerate with  one electron state and $SU_{2}(3)$ point, where two electron states degenerate with one hole state (figure \ref{fig:fig9}(a)). For $\delta = 0.9$ meV nm  four double degeneracy lines (two lines of e-h degeneracy and one line of e-e and one of h-h degeneration) meet in one point SU(4) (figure \ref{fig:fig9}(b)). Increasing  SO interaction further $\delta = 0.8$ meV nm results in the reappearance of  the two SU(3) points again of similar character, but now $SU_{1}(3)$ and $SU_{2}(3)$ change their relative position on magnetic field-gate voltage plane (figure \ref{fig:fig9}(c)). For small values of SO interaction (e.g. $\delta = 0.4$ meV nm) no threefold degeneracy point is observed (figure \ref{fig:fig9}(d)), there occurs only double electron-hole  degeneracy line for this strength of SO interaction. Figure \ref{fig:fig10} and \ref{fig:fig9}(a) illustrate modifications of the ground state diagram with the change of the orientation of magnetic field.  For transverse field electron-hole SU(4) line is visible (figure \ref{fig:fig10}(b)), for $\theta = 89^{\circ}$ two SU(3) points are observed in addition to SU(2) lines (figure \ref{fig:fig9}(a)) and for smaller angel $\theta = 86^{\circ}$ only single SU(3) point is left at the crossing of double degeneracy lines (figure \ref{fig:fig10}(a)).

In the case of strong coupling of the dot with the electrodes, the mentioned degenerations of the electron and hole states make possible the formation of Kondo resonances in which both  electron and hole states participate. For simplicity of numerical analysis  we restrict in our discussion  to the subspace of only two lowest electron states and two highest hole states, what as it is seen from figures \ref{fig:fig8}(c), (d) is justified in the field range where the states degenerate, because other states are distant on the energy scale. The considered regions of energies and fields are shown as grey boxes. The introduced  restriction  considerably simplifies the SBMFA calculations. Similarly to the cases discussed so far,  many-body processes can be described by 16 slave boson operators. Figures \ref{fig:fig11}(a), (b), (c) present partial conductances of CNTQD(33,30) for magnetic field directed at an angle $\theta=89^{\circ}$ to the nanotube axis.   According to  the ground state  diagram presented earlier (figure \ref{fig:fig9}(a)) with double degeneracy lines and two threefold degeneracy points one can expect Kondo SU(2) lines and two different Kondo SU(3) resonances. Vertical dashed lines  on figure \ref{fig:fig9}(a) indicate the cross-sections for which we present  conduction curves. Figure \ref{fig:fig11}(a) presents field dependence of partial conductances along the cross-section  through  SU(3) Kondo state ($SU_{1}(3)$)  with fluctuating states   $|h1+\rangle$, $|h-1+\rangle$ and $|e-1-\rangle$. Figure \ref{fig:fig11}(b)  shows  field dependences of conductances  through  SU(3) Kondo state ($SU_{2}(3)$) involving $|e-1-\rangle$, $|e1-\rangle$  and  $|h-1+\rangle$ states. Figure \ref{fig:fig11}(c) in turn presents  conductance for  a cross-sections through two SU(2) points: hole Kondo state $SU_{1}(2)$ ($|h-1+\rangle$,$|h1+\rangle$) and electron-hole Kondo state $SU_{2}(2)$ ($|h1+\rangle$,$|e-1-\rangle$). In the SU(3) Kondo points partial conductances corresponding to the states taking part in  effective Kondo fluctuations reach value $3/4 (e^{2}/h)$ and the contribution of the fourth channel is negligible. In  SU(2) points two of partial conductances take unitary limit  $e^{2}/h$. Figure \ref{fig:fig11}(d) shows partial conductances of SU(4) Kondo effect occurring for transverse magnetic field. They take the values  $1/2 (e^{2}/h)$ each. Unlike the previously discussed SU(4) Kondo effect, the SU(4) Kondo resonance  appears here for  finite magnetic field. The difference is that the states involved in the processes under discussion do not belong now to the same shell, as in the cases previously analyzed, but two of them are electron states  {$|e-1-\rangle$,$|e1-\rangle$} and two are the  hole states {$|h1+\rangle$,$|h-1+\rangle$}.

\section{Conclusions}

In this paper, we considered the effects of strong correlations in quantum dots formed in carbon nanotubes with small energy gaps. These narrow gaps are formed in otherwise metallic nanotubes by curvature and can be modified by strain or twists. As a result of non-linear dependence of dot energies on the field, restoration  of degeneration is observed for fields dependent on   the value of the atomic potential, controlled by the gate voltage and on the  strength of SO interaction. Lines of degeneracy occur in all Coulomb valleys. There are also threefold  degeneration points in a finite field and fourfold for  zero magnetic field. The resonances of the spin SU(2)  Kondo effect  are characterized by a non-zero orbital moment (quenched spin magnetic  moment) and orbital Kondo resonances exhibit non-zero spin magnetic moment. Kondo SU(3) resonances have a non-zero orbital and spin moments, and in Kondo SU(4) state  both moments are quenched. By changing the value of the energy gap by stress, one can move high symmetry points between different  Coulomb valleys. The SU(4) point  occurs  for zero field. If it appears in the double-occupied region, it separates the SU(3) lines such, that in different parts of the line  there are different SU(3)  Kondo resonances associated with other sets of the dot states.  In a quantum dot formed in a narrow gap nanotube, the electron and hole levels are energetically close enough that some of them  can degenerate  in  magnetic field, which opens  the possibility of Kondo effects of various symmetries in which both electron and hole states  participate. SU(3) points  appear for fields close to the perpendicular orientation of the field with respect to the nanotube axis, and the electron-hole SU(4) Kondo effect is induced in the perpendicular field.\newline

\appendix
\section{}
Slave boson expressions for the charge and spin-orbital fluctuations:
\begin{eqnarray}
&&\langle\langle Q^{2}\rangle\rangle=\sum_{ls}p_{ls}^{2} +4\sum_{lss'}(d_{l}^{2}+d_{ss'}^{2})+9\sum_{ls}t_{ls}^{2}+16f^{2}\nonumber\\&&-(\sum_{ls}p_{ls}^{2} +4\sum_{lss'}(d_{l}^{2}+d_{ss'}^{2})+9\sum_{ls}t_{ls}^{2}+ 16f^{2})^2\nonumber\\
%\nonumber\\
%\nonumber\\
&&\langle\langle Q_{1,-1\uparrow}^{2}\rangle\rangle =
\sum_{s}p_{1s}^2+p_{-1\uparrow}^{2}+4(d_{1}^{2}+d_{\uparrow\uparrow}^{2}+d_{\downarrow\uparrow}^{2})
\nonumber\\&&+d_{\uparrow\downarrow}^{2}+d_{\downarrow\downarrow}^{2}+d_{-1}^{2}+9t_{1\uparrow}^{2}
+4(t_{1\downarrow}^{2}+\sum_{s}t_{-1s}^{2})+9f^{2}-\nonumber\\
&&(\sum_{s}p_{1s}^2+p_{-1\uparrow}^{2}+2(d_{1}^{2}+d_{\uparrow\uparrow}^{2}+d_{\downarrow\uparrow}^{2})
+d_{\uparrow\downarrow}^{2}+d_{\downarrow\downarrow}^{2}+\nonumber\\&&d_{-1}^{2}+
3t_{1\uparrow}^{2}+2(t_{1\downarrow}^{2}+\sum_{s}t_{-1s}^{2})+3f^{2})^{2}\nonumber\\
%\nonumber\\
%\nonumber\\
&&\langle\langle Q_{1\downarrow,-1}^{2}\rangle\rangle =\sum_{s}p_{-1s}^2+p_{1\downarrow}^{2}+4(d_{-1}^{2}+d_{\downarrow\downarrow}^{2}+d_{\downarrow\uparrow}^{2})
\nonumber\\&&+d_{1}^{2}+d_{\uparrow\uparrow}^{2}+
d_{\uparrow\downarrow}^{2}+9t_{-1\downarrow}^{2}
+4(t_{-1\uparrow}^{2}+\sum_{s}t_{1s}^{2})+9f^{2}\nonumber\\&&-
(\sum_{s}p_{-1s}^2+p_{1\downarrow}^{2}+
2(d_{-1}^{2}+d_{\downarrow\downarrow}^{2}+d_{\downarrow\uparrow}^{2})
+d_{1}^{2}+d_{\uparrow\uparrow}^{2}\nonumber\\&&+d_{\uparrow\downarrow}^{2}+3t_{-1\downarrow}^{2}
+2(t_{-1\uparrow}^{2}+\sum_{s}t_{1s}^{2})+3f^{2})^{2}
\end{eqnarray}

\def\refname{References}
%\section*{References}

\end{document}